\documentclass[aps,pra,onecolumn,notitlepage,floatfix,superscriptaddress,nofootinbib]{revtex4-1}
\usepackage{amsmath,amssymb,amsfonts,bbm,graphicx,color}
\usepackage{soul}
\usepackage[utf8]{inputenc}
\usepackage[dvipsnames]{xcolor}
\usepackage[T1]{fontenc}
\usepackage{physics}
\usepackage{times}
\usepackage{amsthm}
\usepackage[caption=false]{subfig}

\usepackage{enumitem}   

\begin{document}

\title{Bath-induced collective phenomena on superconducting qubits: synchronization, subradiance, and entanglement generation}

\author{Marco Cattaneo}
\affiliation{Instituto de F\'{i}sica Interdisciplinar y Sistemas Complejos IFISC (CSIC-UIB),
Campus Universitat Illes Balears, E-07122 Palma de Mallorca, Spain}
\affiliation{QTF Centre of Excellence,  
Department of Physics, University of Helsinki, P.O. Box 43, FI-00014 Helsinki, Finland}
\author{Gian Luca Giorgi}
\affiliation{Instituto de F\'{i}sica Interdisciplinar y Sistemas Complejos IFISC (CSIC-UIB),
Campus Universitat Illes Balears, E-07122 Palma de Mallorca, Spain}
\author{Sabrina Maniscalco}
\affiliation{QTF Centre of Excellence,  
Department of Physics, University of Helsinki, P.O. Box 43, FI-00014 Helsinki, Finland}
\affiliation{QTF Centre of Excellence, Department of Applied Physics, School of 
Science, Aalto University, FI-00076 Aalto, Finland}
\author{Gheorghe Sorin Paraoanu}
\affiliation{QTF Centre of Excellence, Department of Applied Physics, School of 
Science, Aalto University, FI-00076 Aalto, Finland}
\author{Roberta Zambrini}
\affiliation{Instituto de F\'{i}sica Interdisciplinar y Sistemas Complejos IFISC (CSIC-UIB),
Campus Universitat Illes Balears, E-07122 Palma de Mallorca, Spain}


\date{\today}

\begin{abstract}
A common environment acting on a pair of qubits 
gives rise to a 
plethora of different phenomena, such as the
generation of qubit-qubit entanglement, quantum synchronization and subradiance. 
Here we define
time-independent figures of merit for entanglement generation, quantum synchronization and subradiance, and perform an extensive analytical and numerical study of their dependence on model parameters. We also address a recently proposed measure of the 
collectiveness of the dynamics driven by the bath, and find that it almost perfectly witnesses the behavior of entanglement generation. Our results show that synchronization and subradiance can be employed as reliable local signatures 
of an entangling common-bath in a general scenario.
Finally, we propose an experimental implementation of the model based on two transmon qubits 
capacitively coupled to a common resistor, which provides
a versatile quantum simulation platform of the open system in any regime.
\end{abstract}

\maketitle

\section{Introduction}

Common baths inducing dissipation on more than one qubit at a time are known to give rise to different collective phenomena \cite{BreuerPetruccione}. 
Much studied has been superradiance, i.e. the enhancement of the emission power of atoms jointly dissipating into a common environment \cite{Gross1982}.  However, the physics of this system is even richer, with the emergence of additional effects such as
subradiance -- a complementary effect of superradiance \cite{Crubellier1985,Bienaime2012}, qubit-qubit entanglement generation \cite{Braun2002}, and quantum synchronization \cite{Galve2016}. 
The goal of this paper is to map out the parameter regimes of these collective phenomena and to characterize them by time-independent figures of merit. Specifically, the questions we want to address are the following: is there a one-to-one relation between entanglement generation, quantum synchronization and subradiance between a pair of qubits? Is it possible to find a regime where the bath induces one of the above collective phenomenon, but is not able to generate a different one? If we know that, for instance, by tuning a certain model parameter we enhance 
the generation of quantum synchronization and/or subradiance, can we say the same for entanglement? If the answer to the latter question is affirmative, then we may employ quantum synchronization and subradiance as local signatures for the presence of an entangling bath, without
the  resource-intensive complete state tomography.
Mathematically speaking, we aim to investigate the collective features of the  quantum map (dynamical semigroup) describing the open dynamics, or, equivalently, of the corresponding Markovian master equation \cite{BreuerPetruccione}. We  provide specific experimental prescriptions for measuring
each of these figures of merit: one would need to 
start from different initial states, 
e.g. Bell singlet state for subradiance, maximally overlapped state for entanglement, etc.


Note that in addition to quantum synchronization and entanglement generation, we have chosen to study subradiance instead of superradiance between a pair of qubits. Despite its fragility against local noise, subradiance, which manifests  itself through the emergence of a slowly-decaying collective mode in the atomic emission, has been reported in different experiments, see for instance  Refs.~\cite{Pavolini1985,DeVoe1996,Guerin2016,Bhatti2018}. Furthermore, while the presence of a slowly-decaying collective mode is a patent manifestation of subradiance, in the case of only
two qubits superradiance does not exhibit a clear enhancement of the emission power  \cite{Gross1982}. Therefore, a figure of merit for subradiance is more suitable for the quantitative and comparative study we will carry on in this work. Finally, we will also compare the results with a measure of the ``collectiveness'' of the dynamics \cite{Rivas2015} in order to understand whether or not there is a clear, monotonic relation between the strength of the above-mentioned collective phenomena and the spatial correlations induced by the bath. This measure has been tested in a recent experiment \cite{Postler2018}. 

In order to apply our findings to a physical platform of interest for quantum technologies, throughout the paper we will consider a model of two superconducting qubits and, in particular, we will propose to test our predictions through an experimental implementation involving transmon qubits and resistors. Superconducting qubits such as transmons are among the best candidates in the race for the creation of a quantum computer \cite{devoret2013superconducting,Paraoanu2014,wendin2017quantum,arute2019quantum} and configurations such as the ones discussed here
represent a versatile simulation platform for quantum thermodynamics \cite{ronzani2018tunable,Sina2020,Silveri_2017}.
 Within this challenge, one of the major 
 issues is the unavoidable presence of noise \cite{Paladino2002,wendin2017quantum,martinis2014ucsb,Girvin2014,Burnett2019}, which induces dissipation and decoherence and breaks all the most fundamental quantum features, such as superposition or entanglement. Here the study of {\textit{correlated dissipative noise}} induced by a common bath, which is the topic of our paper, is of particular importance. This type of noise appears inevitably due to the coupling to the electromagnetic modes of the nearby waveguides, resonators, and other circuit elements \cite{Koch2007a,Lalumiere2013,Cattaneo2021}. Its investigation has being receiving more and more interest in the recent past; for example, the two-qubit spectroscopy of spatiotemporally correlated noise between two superconducting qubits has been recently proposed \cite{VonLupke2019}. The interest in correlated noise is motivated e.g. by its detrimental effects on the performance of quantum error correction protocols \cite{Devitt2013}, which must be correspondingly modified \cite{Duan1999,Clemens2004,Klesse2005,Novais2006,Aharonov2006,Hayashi2007,Shabani2008,Novais2008,Cafaro2010,
Cafaro2011,Preskill2013,Novais2013,Jacobsen2014,Rivas2015,Layden2019}. A key factor that
for this purpose needs to be taken into account is the generation of qubit-qubit entanglement due to the action of the common bath during the evolution \cite{DeChiara2004,DeChiara2005,Shabani2008}. The use of synchronization and/or subradiance as local signatures of entanglement generation may represent a useful tool for the analysis of these scenarios.

Following this proposal, our findings suggest a novel application 
of quantum synchronization for quantum tasks. 
Recently, transient synchronization was reported to allow for probing
the spectral density of a single qubit 
immersed in a dissipative environment \cite{Giorgi2016}. 
Different forms of synchronization have also
been reported as  beneficial for atomic clocks operation \cite{Gibble2010,Xu2014}.
Finally, the superconducting platform we will discuss could also be implemented as one of the first experimental
realizations of quantum synchronization \cite{Weiner2017,Laskar2019,Koppenhofer2019a}.

The paper is structured as follows: Sec.~\ref{sec:model} presents the model and discusses the system dynamics starting from the master equation of the two qubits, while we introduce and analyse the measures associated to each phenomenon in Sec.~\ref{sec:figuresOfM}. We discuss the results in Sec.~\ref{sec:results}, addressing some analytical limits in Sec.~\ref{sec:anL}, and drawing a numerical comparison between the phenomena for several scenarios in Sec.~\ref{sec:comparison}. We propose the superconducting experimental implementation of our model in Sec.~\ref{sec:experiment}, where we also provide the simulation of the system dynamics in a concrete case. Finally, we draw some concluding remarks in Sec.~\ref{sec:conclusions}.
 
\section{Model}
\label{sec:model}
In order to characterize the collective dissipative effects, we consider two  superconducting transmon qubits \cite{Koch2007a} embedded in a common bath. Specifically, in the experimental proposal of Sec.~\ref{sec:experiment} the bath will be a resistor jointly coupled to both transmon qubits. Such a resistor can be modeled as an infinite collection of bosonic modes, each of them corresponding to an independent fictitious LC circuit, which induce dissipation as described by the  Caldeira-Leggett model \cite{Vool2017}. Remarkably, the circuit Hamiltonian of such a system depends only on the circuit topology, i.e. on how circuit elements are connected between each other, and, for instance, the qubit-qubit distance does not play any role. In other schemes, the role of a common bath may be played by an external electromagnetic environment whose wavelength is larger than the qubit-qubit distance \cite{Dicke1954,Koch2007a}, by a structured phononic environment \cite{Galve2017} or by a resonator to which both qubits are dispersively coupled \cite{VonLupke2019}. The qubits frequencies can be different, reflecting either the inherent variability due to fabrication imperfections or the intentional detuning achieved by applying bias magnetic fields.  In order to identify the effects induced by 
a collective environment, we focus on a set-up in which the qubits are not directly interacting, as this could mask the collective origin of dissipative effects. The Hamiltonian of the overall system reads:
\begin{equation}
\label{eqn:HamiltonianTotal}
\begin{split}
H=& H_S+\sum_k \hbar\Omega_k b_k^\dagger b_k+\mu\sum_k (g_1\sigma_1^x+g_2\sigma_2^x)f_k(b_k+b_k^\dagger),
\end{split}
\end{equation}
where we have introduced the system Hamiltonian $H_S=\frac{\hbar\omega_1}{2}\sigma_1^z+\frac{\hbar\omega_2}{2}\sigma_2^z$, with eigenvectors $\ket{gg},\ket{ge},\ket{eg},\ket{ee}$. Here,  $\ket{e}$ and $\ket{g}$ are respectively the excited and the ground state of a qubit, $\omega_1$ and $\omega_2$ are the frequencies of respectively the first and second qubit, while the dimensionless coefficients $g_1$ and $g_2$ express the weights of the dissipative coupling of each qubit. The parameter $\mu$ is the coupling constant in the units of energy, while each $f_k$ is a real dimensionless number representing the strength of the coupling of the $k-$th mode of the bath to the qubits. The numbers $\{f_k\}_k$ define the spectral density of the common bath \cite{BreuerPetruccione} (see Appendix~\ref{sec:coefficients}), which we will take as Ohmic. This choice is well-justified if the bath is a common resistor, as in the experimental proposal of Sec.~\ref{sec:experiment}: the properties of resistor-induced noise in circuit QED are well-understood and, for instance, the spectral density is known to be Ohmic \cite{ronzani2018tunable,Vool2017}, leading to the correct Johnson-Nyquist spectrum. In other words, the noise generated by a common resistor is the standard thermal noise in the quantum regime \cite{Vool2017}. Finally, the weak-coupling condition reads $\omega_1,\omega_2\gg \mu/\hbar$.

Motivated by the experimental decay observed in transmon qubits (see e.g. 
Ref.~\cite{Burnett2019}), we assume that the system follows a Born-Markov master 
equation \cite{BreuerPetruccione}, and also allow for an additional 
phenomenological local bath on each qubit, inducing local dissipation 
characterized by the time $T_1$.
More details can be found in 
Appendix~\ref{sec:coefficients}. With these prescriptions,  the state of the 
two-qubit system $\rho_S$ obeys the following  master equation under the partial secular approximation \cite{Cattaneo2019}
\begin{equation}
\begin{split}
\label{eqn:masterEqLiouvillian}
\frac{d}{dt}\rho_S(t)=&\mathcal{L}[\rho_S(t)]=-\frac{i}{\hbar}[H_S+H_{LS},\rho_S(t)]+\mathcal{D}[\rho_S(t)],
\end{split}
\end{equation}
where $H_{LS}$ is the Lamb-shift Hamiltonian which reads
\begin{equation}
\label{eqn:lambShift}
H_{LS}=\sum_{j,k=1,2} \hbar(s_{jk}^\downarrow\sigma_k^+\sigma_j^-+{s}_{jk}^\uparrow\sigma_k^-\sigma_j^+),
\end{equation}
while $\mathcal{D}$ is the \textit{dissipator} defined as:
\begin{equation}
\label{eqn:dissipator}
\begin{split}
\mathcal{D}[\rho_S]=&\sum_{j,k=1,2} \gamma_{jk}^\downarrow\left(\sigma_j^-\rho_S\sigma_k^+-\frac{1}{2}\{\rho_S,\sigma_k^+\sigma_j^-\}  \right)+\sum_{j,k=1,2} \gamma_{jk}^\uparrow\left(\sigma_j^+\rho_S\sigma_k^--\frac{1}{2}\{\rho_S,\sigma_k^-\sigma_j^+\}  \right).\\
\end{split}
\end{equation}
The master equation~\eqref{eqn:masterEqLiouvillian} is valid under the weak-coupling assumption $\omega_1,\omega_2\gg \mu/\hbar$ and under the assumption of Markovian bath, which is in general fulfilled when the spectral density is Ohmic \cite{BreuerPetruccione,Cattaneo2019}. Note that we have employed a \textit{partial} secular approximation instead of a full one \cite{Cattaneo2019}, i.e. we keep slowly rotating terms driven by the frequency detuning $\omega_1-\omega_2$ in the derivation of Eq.~\eqref{eqn:masterEqLiouvillian}. This allows us to treat scenarios with very small detuning, where the standard master equation with full secular approximation fails (for further details we refer the reader to Ref.~\cite{Cattaneo2019}).
The coefficients of the master equation~\eqref{eqn:masterEqLiouvillian} contain both a contribution coming from the common bath with Ohmic spectral density, 
introduced in Eq.~\eqref{eqn:HamiltonianTotal}, and a contribution proportional to $1/T_1$, i.e. 
due to the phenomenological local bath acting on each qubit. 
Their form can be found in Appendix~\ref{sec:coefficients}.

The solution of the master equation is obtained by finding the eigenvalues together with the right and the left eigenvectors 
of the Liouvillian $\mathcal{L}$. As discussed in Appendix~\ref{sec:Liouvillian}, the master equation is in partial secular approximation and therefore displays the \textit{phase-covariance symmetry} on the superoperator level \cite{Cattaneo2020}. This means that the Liouvillian superoperator commutes with the \textit{number superoperator}, defined as $\mathcal{N}=[N,\cdot\,]$, where $N$ counts the total number of qubit excitations. Thanks to this symmetry, we can divide the Liouvillian superoperator into five blocks \cite{Cattaneo2020}, which we label using the subscript $d$ such that $d=-2,-1,0,1,2$: $\mathcal{L}=\bigoplus_{d=-2}^2 \mathcal{L}_d$. The blocks have respectively dimension $1,4\times4,6\times6,4\times4,1$, and two of them can be trivially obtained from the others: $\mathcal{L}_{-1}=\mathcal{L}_1^*$, $\mathcal{L}_{-2}=\mathcal{L}_2^*$ (see Appendix~\ref{sec:Liouvillian} for details). We name the eigenvalues of the block $\mathcal{L}_d$ as $\{\lambda_j^{(d)}\}_{j=1}^{\sqrt{dim(\mathcal{L}_d)}}$, the right eigenvectors as $\{\tau_j^{(d)}\}_{j=1}^{\sqrt{dim(\mathcal{L}_d)}}$ and the left eigenvectors as $\{\tilde{\tau}_j^{(d)}\}_{j=1}^{\sqrt{dim(\mathcal{L}_d)}}$. The latter are normalized such that $\Tr[(\tilde{\tau}_j^{(d)})^\dagger\tau_k^{(d)}]=\delta_{jk}$, where $\langle O_1,O_2\rangle=\Tr[O_1^\dagger O_2]$ is the Hilbert-Schmidt inner product of the space of operators acting on the system (see e.g. Ref.~\cite{Bellomo2017}). Using this notation and the master equation~\eqref{eqn:masterEqLiouvillian}, the state of the system at time $t$ is readily obtained \footnote{This decomposition is valid when the Liouvillian is diagonalizable. Eventual special cases need to be treated separately.}:
\begin{equation}
\label{eqn:evstate}
\rho_S(t)=\bigoplus_{d=-2}^2\sum_{j=1}^{\sqrt{dim(\mathcal{L}_d)}} p_{0j}^{(d)}\tau_j^{(d)}\exp(\lambda_j^{(d)}t),
\end{equation}
where $p_{0j}^{(d)}=\Tr[(\tilde{\tau}_j^{(d)})^\dagger\rho_S(0)]$ defines the initial conditions. Eq.~\eqref{eqn:evstate} represents the \textit{normal modes decomposition} of the open dynamics of the system.

\section{Figures of merit for collective phenomena}
\label{sec:figuresOfM}
Next, we introduce the figures of merit of synchronization, subradiance, entanglement generation and correlations 
induced by the presence of a common bath during the relaxation dynamics of the two qubits, and comment on their experimental characterization.
As anticipated, 
our aim is not to detect the simultaneous appearance of these four phenomena for some common initial conditions, 
but to characterize the capability of the common bath to induce some (in general different) open system dynamics displaying them. Therefore, the proposed figures of merit will not depend on time, but only on the parameters of the master equation~\eqref{eqn:masterEqLiouvillian}.
If we can establish that, for instance, the degree of synchronization witnesses that of bath-induced entanglement, then, in an experiment, the former (locally measured) will be instrumental for the choice of the proper conditions to optimize the entanglement production.

Note that the measures of entanglement and collectiveness will be taken as the supremum over time of some time-local indicator (namely, entanglement negativity and measure of correlations in the dynamics \cite{Rivas2015}). In contrast, the figures of merit for synchronization and subradiance will depend on the timescales during which these phenomena occur, and they will not
be based on a time-local indicator.
Indeed, the four measures aim to 
characterize the overall presence of each phenomenon for a given bath-driven quantum dynamics (i.e. quantum map), and not 
at a specific time of the evolution,
as relevant for our purposes. 

\subsection{Quantum synchronization}
\label{sec:sync}
The emergence of transient synchronized dynamics of local observables has been predicted in different quantum systems in recent years 
\cite{Giorgi2012,Manzano2013,Mari2013,Giorgi2013,Xu2014,Zhu2015,Ameri2015,Cabot2017,Cabot2018,Siwiak-Jaszek2019,Cabot2019a,Karpat2019,Eneriz2019,Eshaqi-Sani2020,Tindall2020}. 
This phenomenon is induced by some forms of dissipation into an environment (see 
for instance Refs.~\cite{Galve2016,Giorgi2019} for a review), and in some instance it can persist
asymptotycally \cite{Manzano2013, Cabot2018, Tindall2020}. 
Other forms of synchronization in the quantum regime have also been explored, such as 
entrainment in presence of an external 
driving  or synchronization among self-sustained oscillators \cite{Zhirov2008,Lee2013,Walter2014}.
Also the latter phenomenon is favored by collective dissipation as recently shown for optomechanical systems 
\cite{Cabot2017}. 
For the purpose of this work
we focus on transient spontaneous synchronization mediated by the bath. In particular, we aim to investigate the synchronization of some peculiar oscillating coherences in the qubit dynamics, namely the ones captured by the mean values of the local observables $\sigma_1^x(t)$ and $\sigma_2^x(t)$. Without any external environment action, each qubit's coherences will oscillate as $\langle\sigma_k^x(t)\rangle=\cos(\omega_k t+\phi_k)$, where $\phi_k$ is a fixed phase that depends on the initial conditions. Therefore, in the non-dissipative scenario each qubit oscillates at a different frequency. In contrast, the presence of dissipation can induce a regime where both qubits oscillate at a synchronized frequency \cite{Giorgi2013}. In particular, under the dynamics driven by Eq.~\eqref{eqn:masterEqLiouvillian}, the mean values of these local observables
 can be completely recovered by analyzing the Liouvillian block $\mathcal{L}_1$ only \cite{Bellomo2017,Cattaneo2020}:
\begin{equation}
\label{eqn:sigmaX}
\begin{split}
\langle\sigma_k^x(t)\rangle=\sum_{j=1}^4 2 \abs{p_{0j}^{(1)} c_{jk}^{(1)}}e^{\Re(\lambda_j^{(1)})t}\cos(\Im(\lambda_j^{(1)})t+\varphi_{jk}^{(1)}),
\end{split}
\end{equation}
with $c_{jk}^{(1)}=\Tr[\sigma_k^x \tau_{j}^{(1)}]$ and $\varphi_{jk}^{(1)}=\arg(p_{0j}^{(1)} c_{jk}^{(1)})$. Here, $\Re(\lambda_j^{(1)})$ and $\Im(\lambda_j^{(1)})$ denote the real and imaginary parts of the eigenvalue $\lambda_j^{(1)}$.

Note that we have chosen to investigate synchronization of the mean values of $\sigma_k^x$. Any other choice of observable detecting oscillating coherences, such as $\sigma_k^y$, would be perfectly possible as well, and would lead to analogous results \cite{Bellomo2017}. For the sake of simplicity, we focus on $\sigma_k^x$ only.

A well-known figure of merit to detect the synchronized dynamics of two observables is the Pearson coefficient \cite{Galve2016}, 
whose definition can be found in Eq.~\eqref{eqn:pearson} of Appendix~\ref{sec:derFig}. 
Being a temporal correlation of the measured coherences dynamics of each qubit, this quantity can be easily experimentally measured. For our theoretical analysis
in different parameter regimes we introduce a 
time-independent figure of merit. This will only depend
on the coefficients of the master equation \eqref{eqn:masterEqLiouvillian}, having fixed some specific initial conditions.
In fact, the emergence of synchronization is due to a  mode  of the block $\mathcal{L}_1$ (let us say the one with eigenvalue $\lambda_{4}^{(1)}$) decaying much slower than any other mode \cite{Giorgi2012,Manzano2013,Giorgi2013}.  According to Eq.~\eqref{eqn:sigmaX}, at a certain time $t_S$, $\sigma_1^x(t)$ and $\sigma_2^x(t)$ will synchronize at frequency $\Im(\lambda_{4}^{(1)})$. A measure of synchronization can be introduced as follows:
\begin{equation}
\label{eqn:syncMeasure}
Syn=\abs{\frac{\log 100}{t_{S}\Re(\lambda_4^{(1)})}}.
\end{equation} 

$Syn$ describes the ratio between the decay time of the slowest-decaying mode 
$\lambda_4^{(1)}$ and the \textit{synchronization time} $t_S$, defined as the time at which the contribution to $\langle\sigma_1^x(t)\rangle$ and $\langle\sigma_2^x(t)\rangle$ of 
this mode
is 100-times bigger than the one of any other mode (see the definition in Eq.~\eqref{eqn:syncTime} of Appendix~\ref{sec:derFig}). Choosing the number 100 in the definition of synchronization time (and correspondingly, in the factor $\log 100$ in Eq.~\eqref{eqn:syncMeasure}) is heuristic. We do so in order to characterize the ``disappearance'' of all modes but the synchronized one, however, other choices are possible. Eventually, what we are characterizing is the trade-off between the dissipation timescale and synchronization timescale, and the qualitative behavior of the figure of merit $Syn$ should not depend on this numerical choice. It is important to choose the same number in Eq.~\eqref{eqn:syncMeasure} and in Eq.~\eqref{eqn:syncTime} (i.e., if we set $\log 50$ in Eq.~\eqref{eqn:syncMeasure} then we will define the synchronization time as the time at which the synchronization mode is 50-times larger than any other one). 
By doing so, $Syn>1$ 
{indicates} that  synchronization appears before the ``relaxation'' of the 
slowest decaying mode. The bigger $Syn$, the more detectable the synchronization 
in the system. We choose $\rho_S(0)=\ket{\psi_{Syn}}\bra{\psi_{Syn}},$ as 
initial state of the evolution, where $\ket{\psi_{Syn}}=(\cos\pi/4 
\ket{e}+\sin\pi/4\ket{g})\otimes (\cos\pi/3 \ket{e}+i\sin\pi/3\ket{g})$,
as such a state has a large amount of initial coherence and at the same time 
is not symmetric under the exchange of the two qubits, which avoids pathological 
behaviors for some choice of parameters due to symmetry. In the case of transmon qubits, this separable state can be prepared by standard techniques using microwave Rabi pulses with frequencies $\omega_1$ and $\omega_2$ applied to to each qubit respectively.

\subsection{Subradiance}

In order to introduce a measure able to capture the appearance of subradiance in our model, we focus on the observables describing the population of the excited state of 
each qubit, defined as: 
\begin{equation}
\label{eqn:emissionObs}
\begin{split}
P_1^e=&\ket{e}\bra{e}\otimes\mathbb{I},\\
P_2^e=&\mathbb{I}\otimes\ket{e}\bra{e}.\\
\end{split}
\end{equation} 
The dynamics of $P_1^e$ and $P_2^e$ can be completely recovered by analyzing the Liouvillian block $\mathcal{L}_0$ only \cite{Bellomo2017,Cattaneo2020}:
\begin{equation}
\label{eqn:sigmaZ}
\begin{split}
\langle P_k^e(t)\rangle=\sum_{j=1}^6  &p_{0j}^{(0)} h_{jk}^{(0)}e^{\lambda_j^{(0)}t},
\end{split}
\end{equation}
where $h_{jk}^{(0)}=\Tr[P_k^e \tau_{j}^{(0)}]$.

As said in the introduction, subradiance entails the presence of a slowly-decaying collective mode in the dynamics of both the observables 
$P_1^e(t)$ and $P_2^e(t)$ defined in Eq.~\eqref{eqn:emissionObs}, which remains alive even when all the other modes have disappeared. $\mathcal{L}_0$ always has at
least one zero eigenvalue, corresponding to the steady state of the dynamics \cite{Cattaneo2020}. We do not consider the latter, and we focus on the remaining five. Analogously to the problem of quantum synchronization, we introduce the measure of subradiance as:
\begin{equation}
\label{eqn:syncSub}
Sub=\abs{\frac{\log 100}{t_{B}\Re(\lambda_5^{(0)})}},
\end{equation}
where $\lambda_5^{(0)}$  is the slowest-decaying mode of $\mathcal{L}_0$ (apart from the non-decaying one) and $t_B$ is the \textit{subradiance time} at which the slowest-decaying component in the mean value of $P_1^e(t_B)$ and $P_2^e(t_B)$ is 100-times bigger than that of any other mode (apart from the steady state one), whose expression can be found in Eq.~\eqref{eqn:subTime} of Appendix~\ref{sec:derFig}. As for the case of synchronization, choosing the number 100 in the above definition is heuristic and other choices are possible (see the discussion in the previous section). We assume to start the evolution in the \textit{subradiant state} $(\ket{eg}-\ket{ge})/\sqrt{2}$, which in a perfectly symmetric scenario ($\omega_1=\omega_2$, $g_1=g_2$ and without local dissipation) lives in a decoherence-free subspace. This state is a Bell state that can be prepared by standard methods (see also Section 5). For example, starting with the qubits in the state $|g\rangle |g\rangle$, we can apply on the first qubit an X gate  ($\sigma_{x}$ with our notations) followed by a Hadamard gate, then use this qubit as a control for a CNOT gate with the second qubit as target. Finally, we apply another $X$ gate on the first qubit.
As $Sub$ describes the ratio between and the decay time of the slowest-decaying mode $\lambda_5^{(0)}$
and the subradiance time,  $Sub>1$ implies that subradiance appears before the ``relaxation'' of the slowest decaying mode. If we consider the symmetrical 
scenario leading to a decoherence-free subspace, we have $Sub=\infty$, since the initial state would be a steady state of the evolution. In general terms, the 
bigger $Sub$ the more detectable the subradiance in the system.

\subsection{Entanglement generation}
\label{sec:entanglement}
The fact that a common bath can entangle two initially uncorrelated quantum systems has been discussed in several scenarios \cite{Plenio1999,Beige2000,Basharov2002,Braun2002,Plenio2002,Kim2002,Benatti2003a,Yi2003,Kraus2004,Lendi2007,Diehl2008,PhysRevB.78.064503,Hor-Meyll2009,Benatti2010,Caruso2010,Gonzalez-Tudela2011,Wang2013,Reiter2013,Leghtas2013} and also investigated on some experimental platforms \cite{Krauter2011,Lin2013,Shankar2013}.
In contrast, for separate baths entanglement may exist only if direct coupling is present and can be amplified by local drives \cite{Li2008,Li2009,LiParaoanu2010}.
Following the works by Benatti et al. on Markovian dynamics \cite{Benatti2003a,Benatti2010,Benatti2008}, we can provide sufficient conditions on the capability of the master equation~\eqref{eqn:masterEqLiouvillian} to generate entanglement for given configurations of the Hamiltonian parameters. The conditions are presented in Appendix~\ref{sec:entProof}.

As a measure of entanglement of the state of the system we choose the \textit{negativity}, which is a well-defined, easily computable entanglement monotone \cite{PhysRevA.65.032314}. We expect any other measure of entanglement, such as entanglement of formation, to produce very similar qualitative results \cite{RevModPhys.81.865}. Negativity is given by
\begin{equation}
\label{eqn:negativity}
\mathcal{N}(t)=\sum_{r_j(t)<0}\abs{r_j(t)},
\end{equation}
where $r_j(t)$ are the eingevalues of the partial transpose of the state of the system $\rho_S(t)$ with respect to the second qubit. $\rho_S(t)$ evolves according to the master equation~\eqref{eqn:masterEqLiouvillian}, therefore the negativity depends on time as well, and we choose as measure of the entangling power of the bath the maximum value over the whole evolution:
\begin{equation}
\label{eqn:negMeasure}
\mathcal{N}_M=\sup_{t>0}[\mathcal{N}(t)].
\end{equation}
This definition clearly depends on the initial conditions. As separable initial state of the evolution, we set $\rho_S(0)=\rho_{C}=1/4\sum_{j,k,l,m=e,g}\ket{jk}\bra{lm}$, which is the maximally coherent state. We choose so by considering that the majority of quantum algorithms start from $\rho_{C}$, given that the latter is obtained by applying a Hadamard gate on the state with all the qubits in $0$ (here $\ket{gg}$) \cite{NielsenChuang}, which on a superconducting platform can be realized by applying resonant microwave Rabi pulses to each qubit. Some different choices of initial state will be discussed in Appendix~\ref{sec:entProof} and Sec.~\ref{sec:initCond}.

The computation of $\mathcal{N}_M$ requires the complete knowledge of the state of the system $\rho_S(t)$ at any time $t$. This can be obtained experimentally by means of the full-state tomography \cite{Filipp2009,Shankar2013}, based on joint measurements on both the qubits at the same time, which is highly resource-intensive.

\subsection{Collectiveness of the dynamics}
To quantify the collectiveness of the dynamics of two qubits driven by a common bath, we employ a measure recently introduced by Rivas and Müller \cite{Rivas2015}. It quantifies how much a given quantum evolution $\mathcal{E}(t)$ of two subparties $1$ and $2$ differs from a totally uncorrelated dynamics written as $\mathcal{E}_1(t)\otimes \mathcal{E}_2(t)$.

The measure is based on the Choi-Jamiołkowski isomorphism which makes use of the state $\ket{\Psi_{SS'}}$, living in $\mathcal{H}_S\otimes\mathcal{H}_{S'}$, where $\mathcal{H}_S$ is the Hilbert space of the system and $\mathcal{H}_{S'}$ a copy of it. $\ket{\Psi_{SS'}}$ is defined as:
\begin{equation}
\label{eqn:choiState}
\ket{\Psi_{SS'}}=\frac{1}{2}\sum_{j,k=e,g}\ket{jk}_{12}\otimes\ket{jk}_{1'2'}.
\end{equation}
The subscripts indicate that the generic single-qubit state $\ket{j}$ refers to the qubit $1$ or $2$ of $\mathcal{H}_S$ or to the qubit $1'$ or $2'$ of $\mathcal{H}_{S'}$.
Let us name $\mathcal{E}(t)=\exp(\mathcal{L}t)$ the quantum map describing the evolution until time $t$. Then, we can introduce a $16\times 16$ matrix $\Phi_\mathcal{E}(t)$, defined by
\begin{equation}
\label{eqn:choiMatrix}
\Phi_\mathcal{E}(t)=\mathcal{E} (t)\otimes\mathbb{I}_{S'}[\ket{\Psi_{SS'}}\bra{\Psi_{SS'}}].
\end{equation}
Since $\mathcal{E}(t)$ is completely positive, $\Phi_\mathcal{E}(t)$ is positive-semidefinite and has trace $1$, i.e. is a state in the Hilbert space $\mathcal{H}_S\otimes\mathcal{H}_{S'}$. The measure of correlations in the dynamics $\bar{I}$ is defined as the normalized quantum mutual information of $\Phi_\mathcal{E}(t)$:
\begin{equation}
\label{eqn:measureCorr}
\begin{split}
\bar{I}(\mathcal{E}(t))=&\frac{I(\Phi_\mathcal{E}(t))}{4\log 2}\\
=&\frac{S(\Tr_{11'}[\Phi_\mathcal{E}(t))])+S(\Tr_{22'}[\Phi_\mathcal{E}(t))])-S(\Phi_\mathcal{E}(t))}{4\log 2},
\end{split}
\end{equation}
where $S$ is the von Neumann entropy and $\Tr_{11'(22')}$ is the partial trace on the qubit $1$ and $1'$ ($2$ and $2'$). $0\leq \bar{I}(\mathcal{E}(t))\leq 1$, it is null if and only if the dynamics is uncorrelated, and it satisfies a fundamental principle which allows one to consider the correlations in the dynamics as a resource (see Ref.~\cite{Rivas2015} for details), therefore it is a well-defined measure. It can be shown that it reaches the value $\bar{I}(\mathcal{E}(t))=1$ only if $\mathcal{E}(t)$ is unitary.

$\bar{I}(\mathcal{E}(t))$ is a measure for the correlation in the quantum map that evolves the state until time $t$. Since we would like to have a figure of merit independent of time, following the original reference \cite{Rivas2015} we choose:
\begin{equation}
\label{eqn:measureCorrSup}
\bar{I}_M=\sup_{t>0}[\bar{I}(\mathcal{E}(t))].
\end{equation}
For convenience, let us term the measure in Eq.~\eqref{eqn:measureCorrSup} as \textit{collectiveness}. {The experimental computation of $\bar{I}_M$ requires {quantum process tomography} which is highly non-trivial \cite{Postler2018}, and in the present work we consider this figure of merit only as an abstract indicator to be compared with the other measures.}

\section{Results}
\label{sec:results}
Although synchronization, subradiance, entanglement generation and correlations in the dynamics of decoupled qubits share a common origin in this system,
namely the presence of a collective bath, they are essentially four different physical phenomena.
From a mathematical perspective, this is displayed by the block structure of the Liouvillian 
superoperator discussed in Appendix~\ref{sec:Liouvillian}. We observe that quantum synchronization is described by the block $\mathcal{L}_1$ (Eq.~\eqref{eqn:sigmaX}), 
subradiance by the block $\mathcal{L}_0$ (Eq.~\eqref{eqn:sigmaZ}), while both the negativity Eq.~\eqref{eqn:negativity} and the collectiveness Eq.~\eqref{eqn:measureCorr} 
may in general depend on all the blocks of the Liouvillian. We know that, only in the case with zero temperature, a parametric dependence induces the same separation 
between the decay rates characterizing synchronization and subradiance \cite{Bellomo2017}, but in more general scenarios there is no clear relation between the phenomena, 
given that each Liouvillian block has an independent behavior. Therefore, we first discuss the behaviour of the four figures of merit for some simple limits of the model parameters, which can be computed analytically, and then we perform a more general, deep numerical investigation in order to compare the strength of each phenomenon in
different scenarios and their possible concomitance. Note that the discussion on the analytical limits is valid not only for the case of transmon qubits immersed in an Ohmic thermal bath, but for any system of two qubits dissipatively coupled to an environment.

\subsection{Analytical limits}
\label{sec:anL}
\subsubsection{$T=0$}
In the zero temperature case, the form of the eigenvalues of each block of the Liouvillian superoperator is known, as discussed in Appendix~\ref{sec:an}. In particular, we find that the separation between the eigenvalues corresponding to the two slowest modes of both $\mathcal{L}_0$ and $\mathcal{L}_1$ has the same expression, and can be written as (the principal square root is taken):
\begin{equation}
\label{eqn:separation}
\begin{split}
\Delta \lambda =& \lambda_5^{(0)}-\lambda_4^{(0)} =\lambda_4^{(1)}-\lambda_3^{(1)}\\
=&\sqrt{(\gamma_{12}^\downarrow+2i s_{12}^\downarrow)(\gamma_{21}^\downarrow+2i s_{21}^\downarrow)+\left(\frac{\gamma_{11}^\downarrow-\gamma_{22}^\downarrow}{2}+ i\Delta \omega\right)^2},
\end{split}
\end{equation}
with $\Delta\omega=\omega_1-\omega_2$.
$\Re(\Delta \lambda)$ provides the separation between the intensity of the decay rates of the slowest modes, that plays an important role in the definition of $Syn$ and $Sub$ respectively in Eqs.~\eqref{eqn:syncMeasure} and~\eqref{eqn:syncSub}. The case of high detuning is trivial, since it leads to the validity of the full secular approximation \cite{Cattaneo2019}, i.e. the qubit-qubit correlations in the master equation become negligible. As for the time-scales separation, it does not arise since $\Delta \lambda$ is almost purely imaginary. 

More interesting for our purpose is the case of small detuning. We start assuming balanced weights, $g_1=g_2$. Under these assumptions, $\gamma_{11}-\gamma_{22}\approx 0$, therefore we have $\Delta \lambda\approx\sqrt{(\gamma_{12}^\downarrow+i s_{12}^\downarrow)(\gamma_{21}^\downarrow+i s_{21}^\downarrow)-\Delta \omega^2}$. Looking at $\Delta \lambda$ as a function of the detuning, we easily observe that $\frac{d}{d\Delta \omega}\Re(\Delta\lambda)<0$, i.e. the separation between the real part of the eigenvalues of the two slowest eigenmodes decreases as a function of the detuning. According to the discussion in Appendix~\ref{sec:an}, for $g_1=g_2$ and $\omega_1=\omega_2$ the eigenvectors associated to the slowest eigenvalues have the maximum projection over $\sigma_1^x$ and $\sigma_2^x$ for synchronization, and $P_1^e$ and $P_2^e$ for subradiance. Then, we can conclude that, in the balanced case $g_1=g_2$ and for $T=0$, increasing the detuning hinders $Syn$ and $Sub$. Moreover, for $g_1=g_2$ and finite detuning, decreasing the strength of the qubit-bath coupling also hinders $Syn$ and $Sub$, since $\lim_{\mu\rightarrow 0^+}\Delta \lambda \approx \sqrt{-\Delta \omega^2}$, which has zero real part.

As for entanglement and collectiveness, these phenomena depend on the amount of correlations that  dynamically build up among the qubits. Therefore, they will not arise in the limits in which such correlations disappear. In particular, $\mathcal{N}_M=0$ and $\bar{I}_M=0$ (i) for $\mu^2\ll 1/T_1$ (the collective action of the bath is negligible with respect to the local incoherent dissipation), {(ii)} for $g_2\rightarrow 2, g_1\rightarrow 0$ and viceversa (the bath acts only on a single qubit and not on the other), and (iii) for $\Delta\omega =O(\omega_1)$ (the detuning is of the order of the qubit frequencies, and therefore the full secular approximation applies). We observe that entanglement and collectiveness display the same behavior as synchronization and subradiance in these limits. 

\subsubsection{$T\rightarrow\infty$}
In the infinite temperature
limit, exact analytical results can be derived for $g_1=g_2$, $\Delta \omega=0$, so that there is a single dissipative coefficient $\gamma=\gamma_{jk}^\downarrow=\gamma_{jk}^\uparrow$. Under these assumptions, first of all we observe that the measure of subradiance has an infinite value, $Sub=\infty$, since the subradiant state lives in a decoherence-free subspace, i.e. $\lambda_5^{(0)}=0$, and it never decays, independently of the chosen temperature. Besides, we see that by increasing the temperature we speed up all the remaining decaying rates
$\lambda_1^{(0)},\ldots,\lambda_4^{(0)}$ in Eq.~\eqref{eqn:subTime}, and therefore, in general terms we bring forward the subradiance time. For continuity, we expect a similar behavior also when the detuning is not zero anymore (and therefore the subradiance measure has a finite value), that is to say, subradiance is improved for higher temperatures.

In the case of synchronization, we can compute the eigenvalues of the block $\mathcal{L}_1$, whose expressions can be found in Appendix~\ref{sec:an}. We observe that the difference between the two slowest eigenvalues is purely imaginary:
\begin{equation}
\lambda_4^{(1)}-\lambda_3^{(1)}=-2is_+,
\end{equation}   
with $s_+=s_{12}^\downarrow+s_{21}^\uparrow$. Therefore, the real part of the difference is zero and synchronization can never appear, since there will always be at least two modes with different frequency in the dynamics of the observables in Eq.~\eqref{eqn:sigmaX}. Hence, $Syn\rightarrow 0$ for $T\rightarrow \infty$.

Finally, let us address entanglement and collectiveness. For these figures of merit we are not able to derive an analytical solution at any time. However, previous works have shown that the entanglement generated by the action of a common bath vanishes at infinite temperature \cite{Lendi2007,badveli2020dynamics}. Let us understand why. First, note that as we increase the temperature, we strongly increase the dissipative coefficient $\gamma$, proportional to $\coth(\beta\hbar\omega/2)$, while we do not observe such a strong enhancement of the collective Lamb-shift term $s_+$ (see e.g. Eq.~\eqref{eqn:coefficients} in Appendix~\ref{sec:coefficients}). As extensively discussed in the literature, the capability of a common bath to entangle a couple of non-interacting qubits strongly depends on the collective Lamb-shift term \cite{Solenov2007}. This is also suggested by our analysis of the sufficient conditions for an entangling bath presented in Appendix~\ref{sec:entProof}: if $s_+\rightarrow 0$ and $T\rightarrow \infty$, the sufficient condition is not satisfied anymore. Therefore, we can expect that, in the limit $\gamma\gg s_+$, this inequality for the sufficient condition in Appendix~\ref{sec:entProof} is only weakly satisfied, an indication of small entanglement. If this is true, we obtain that $\mathcal{N}_M\rightarrow 0^+$ for $T\rightarrow\infty$. In Appendix~\ref{sec:an} we show that this hypothesis is correct through a careful study of the trade-off between $\gamma$ and $s_+$. Furthermore, we can employ a similar argument for the measure of collectiveness, showing that it decreases (although does not vanish) for $T\rightarrow\infty$: if $\gamma\gg s_+$, the collective Lamb-shift term becomes negligible in the master equation and the collective action of the quantum map derives only from the dissipative part. On the contrary, when the Lamb-shift term is relevant, its action brings an additional collective term to the master equation, increasing the value of $\bar{I}_M$. The analysis in Appendix~\ref{sec:an} confirms this claim.

\begin{figure*}
\centering
\subfloat[]{%
  \includegraphics[scale=0.34]{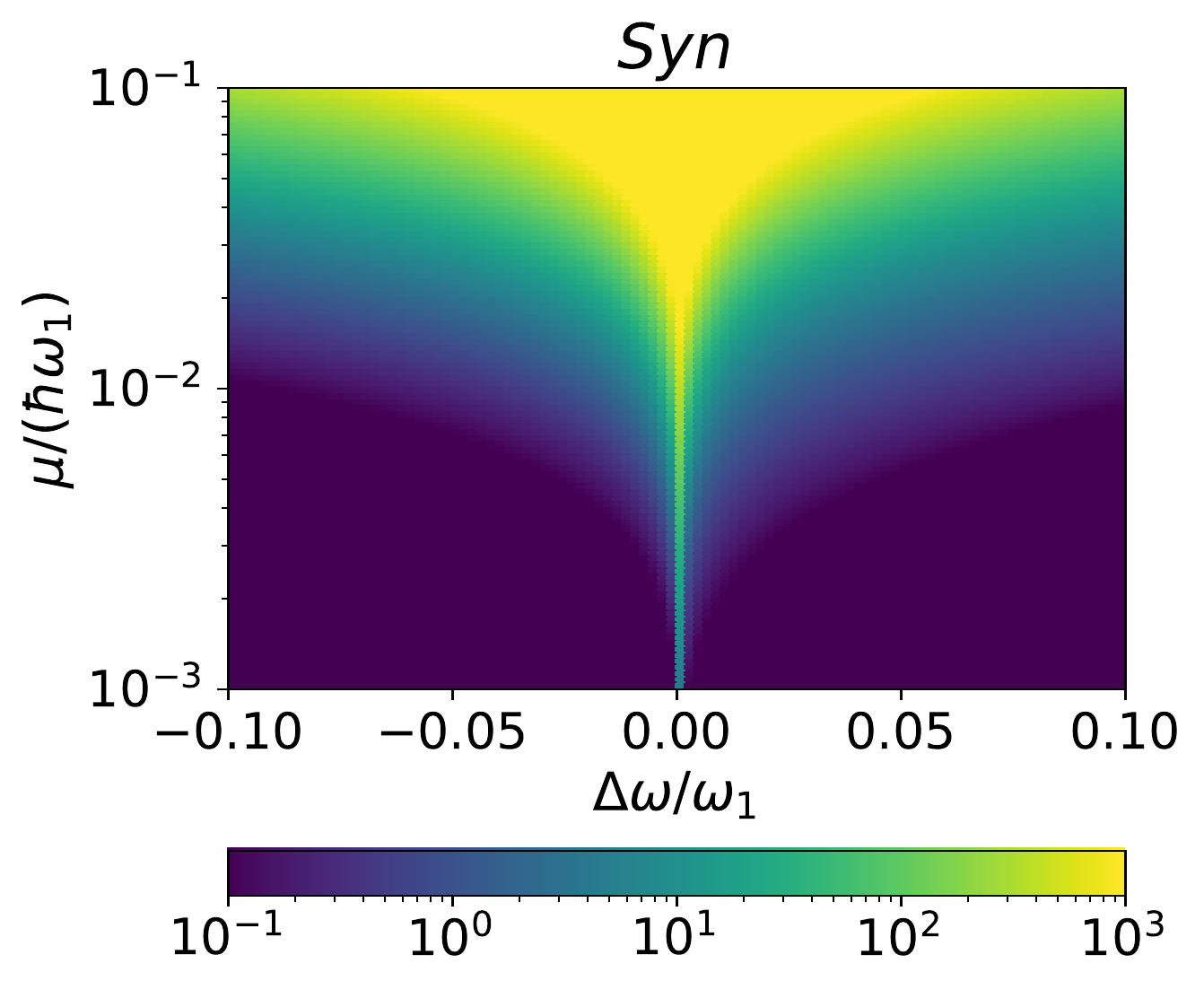}%
}
\subfloat[]{%
  \includegraphics[scale=0.34]{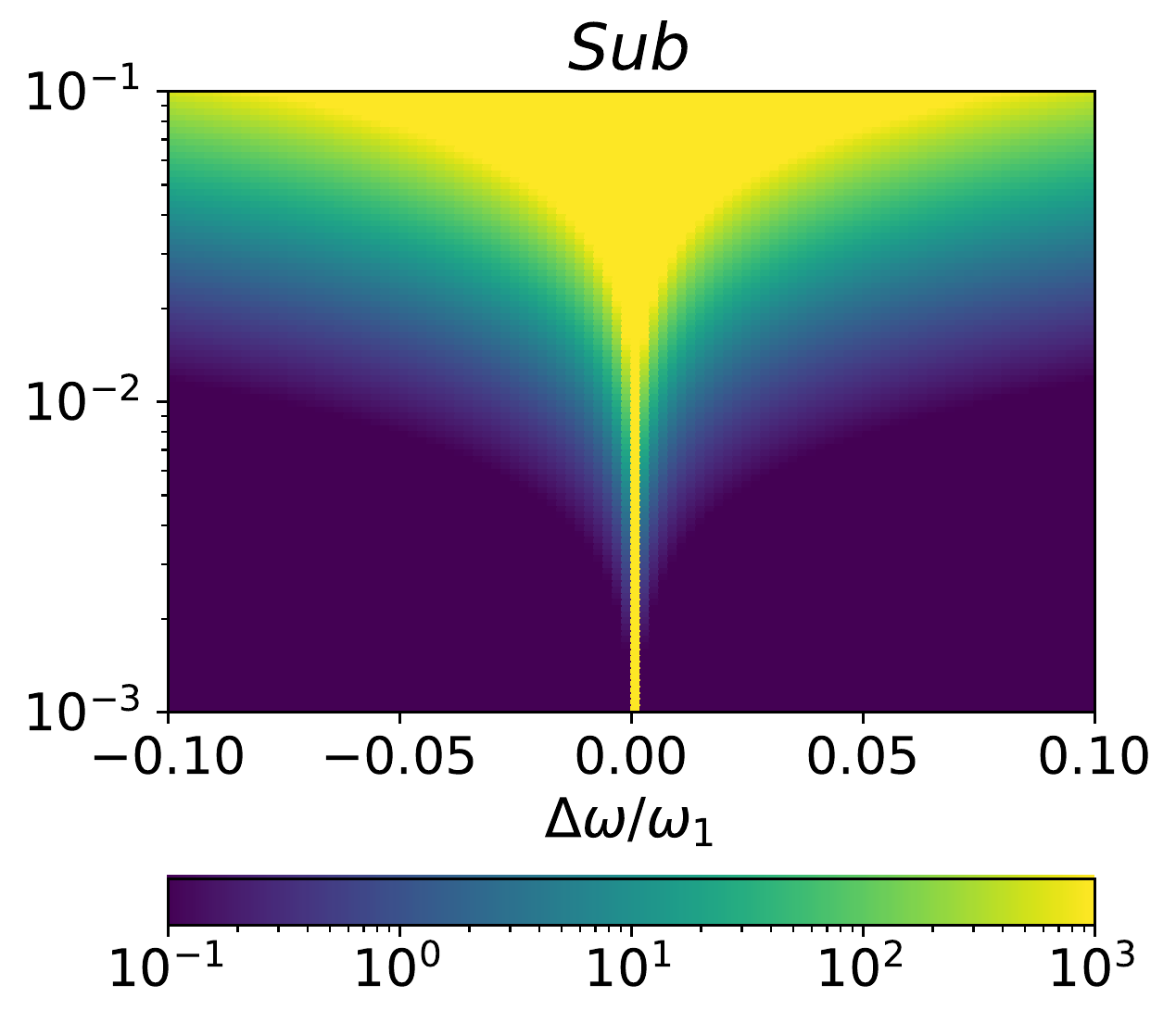}%
}
\subfloat[]{%
  \includegraphics[scale=0.34]{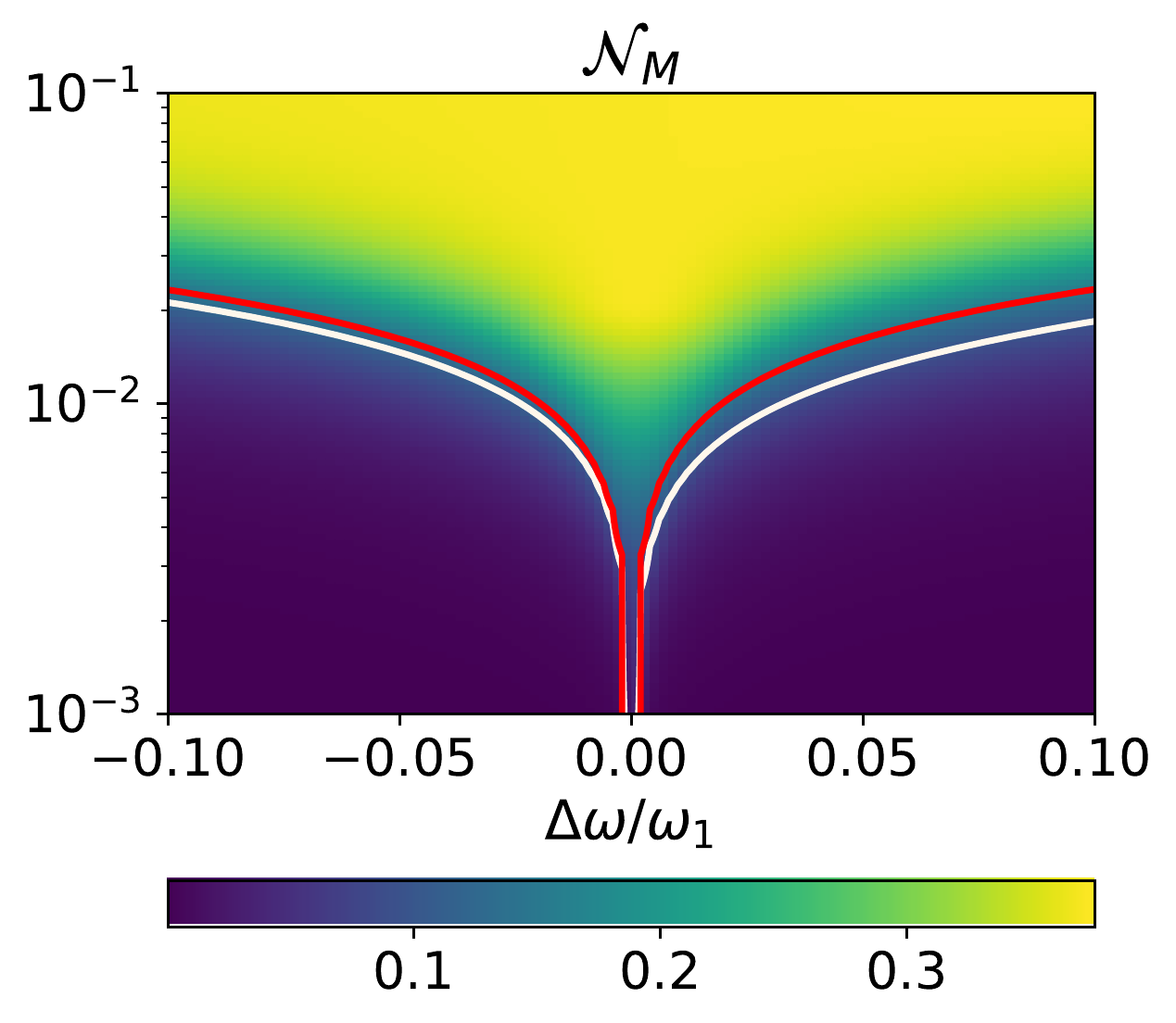}%
}
\subfloat[]{%
  \includegraphics[scale=0.34]{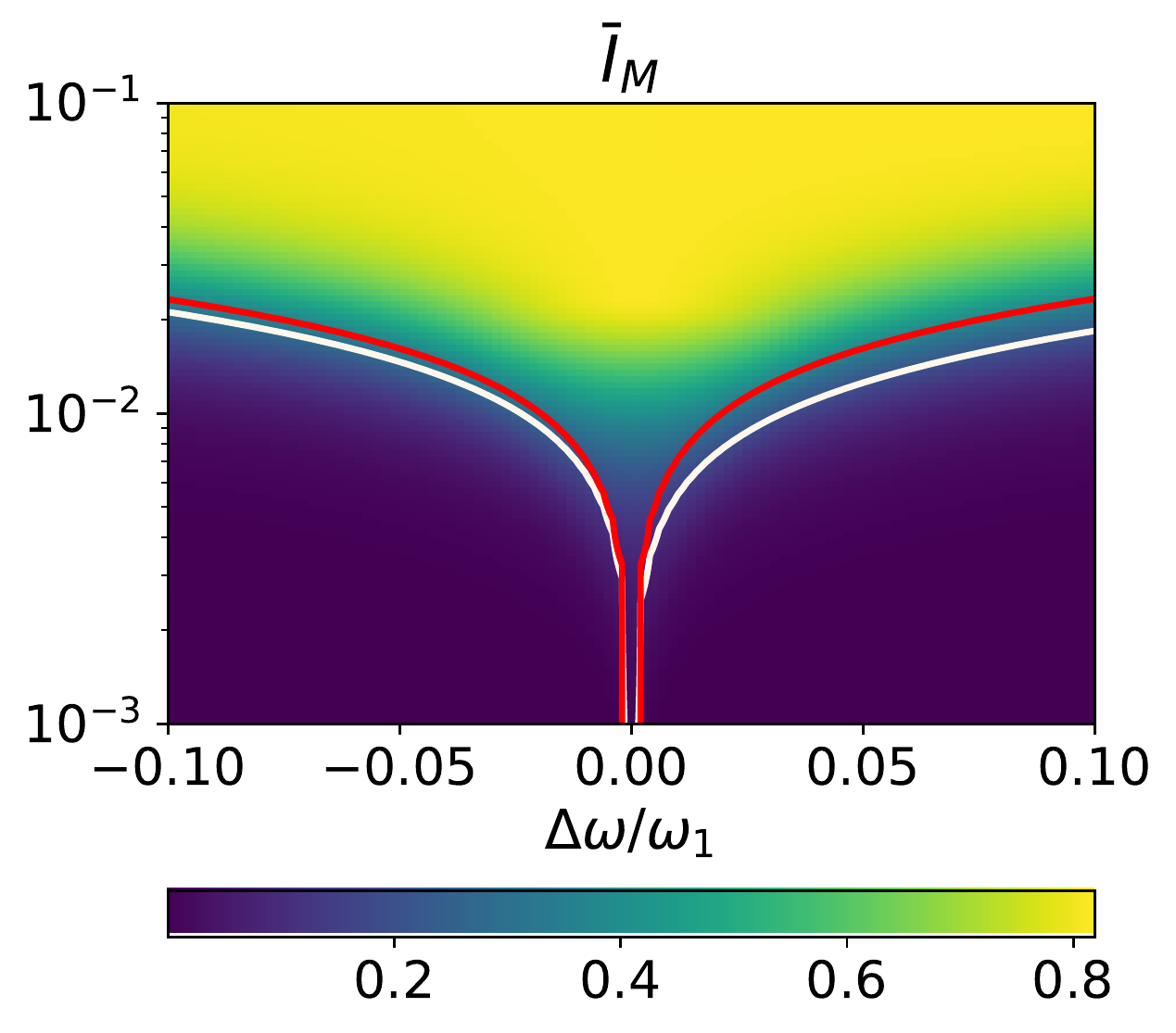}%
}\\
\subfloat[]{%
  \includegraphics[scale=0.34]{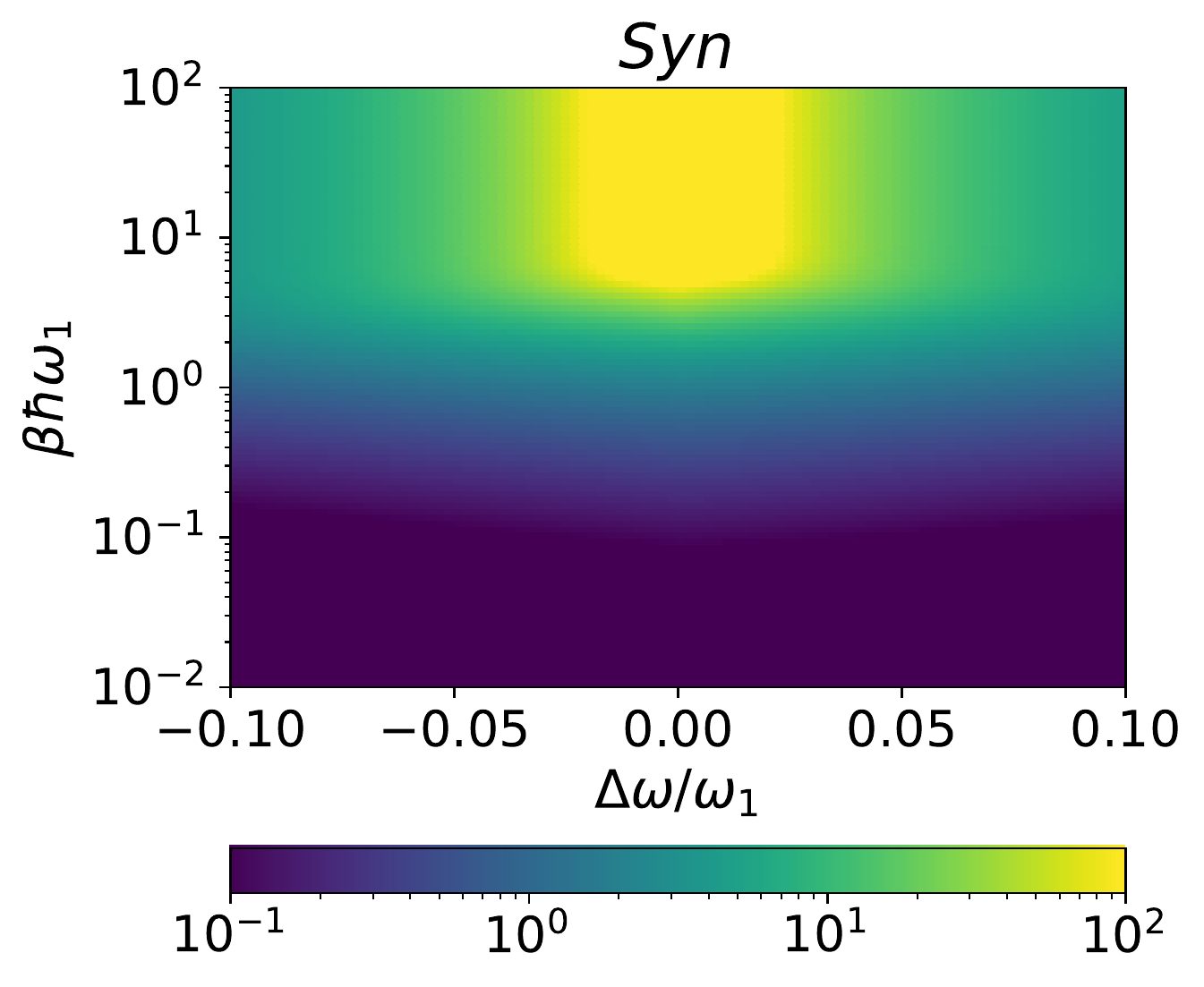}%
}
\subfloat[]{%
  \includegraphics[scale=0.34]{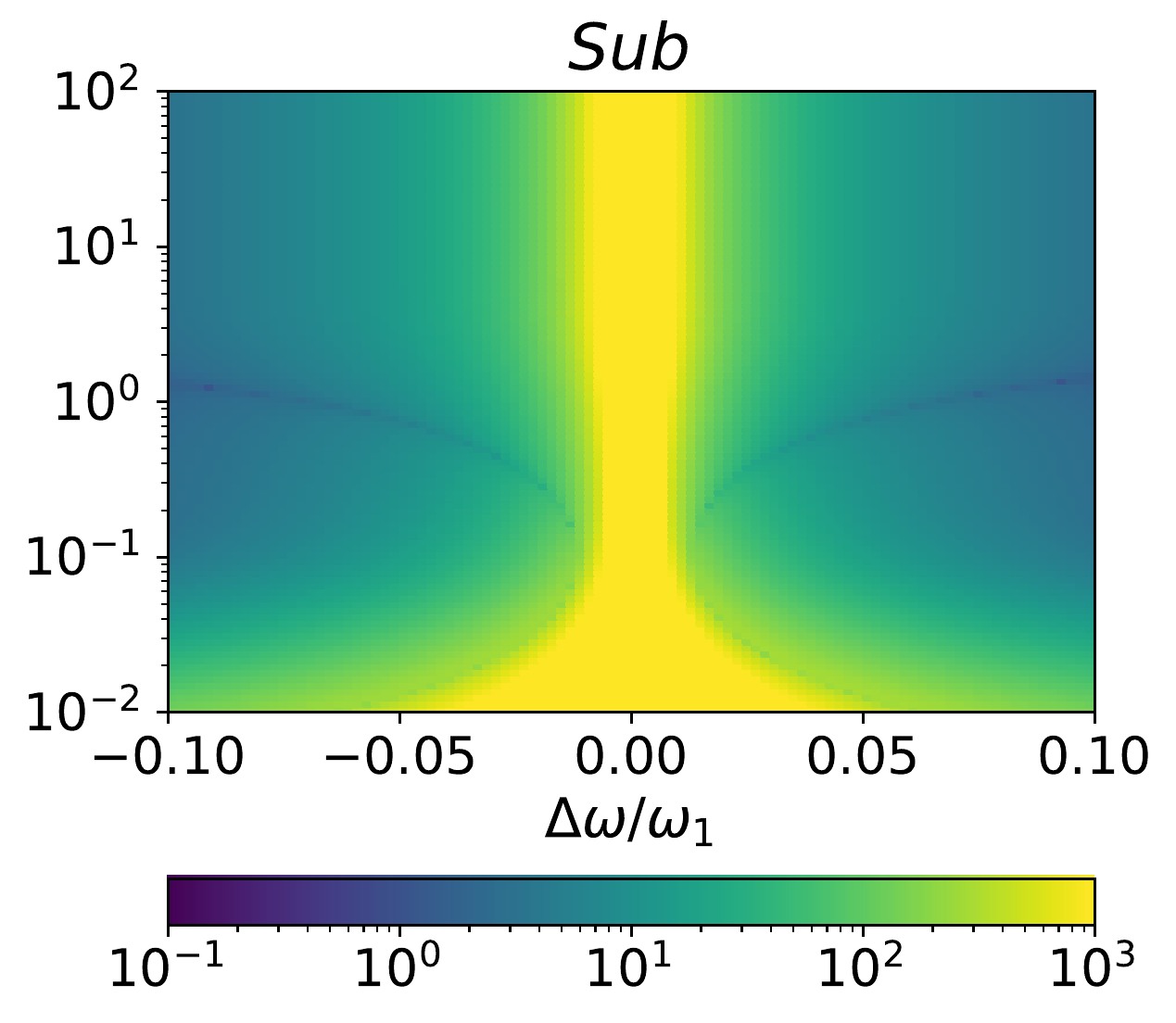}%
}
\subfloat[]{%
  \includegraphics[scale=0.34]{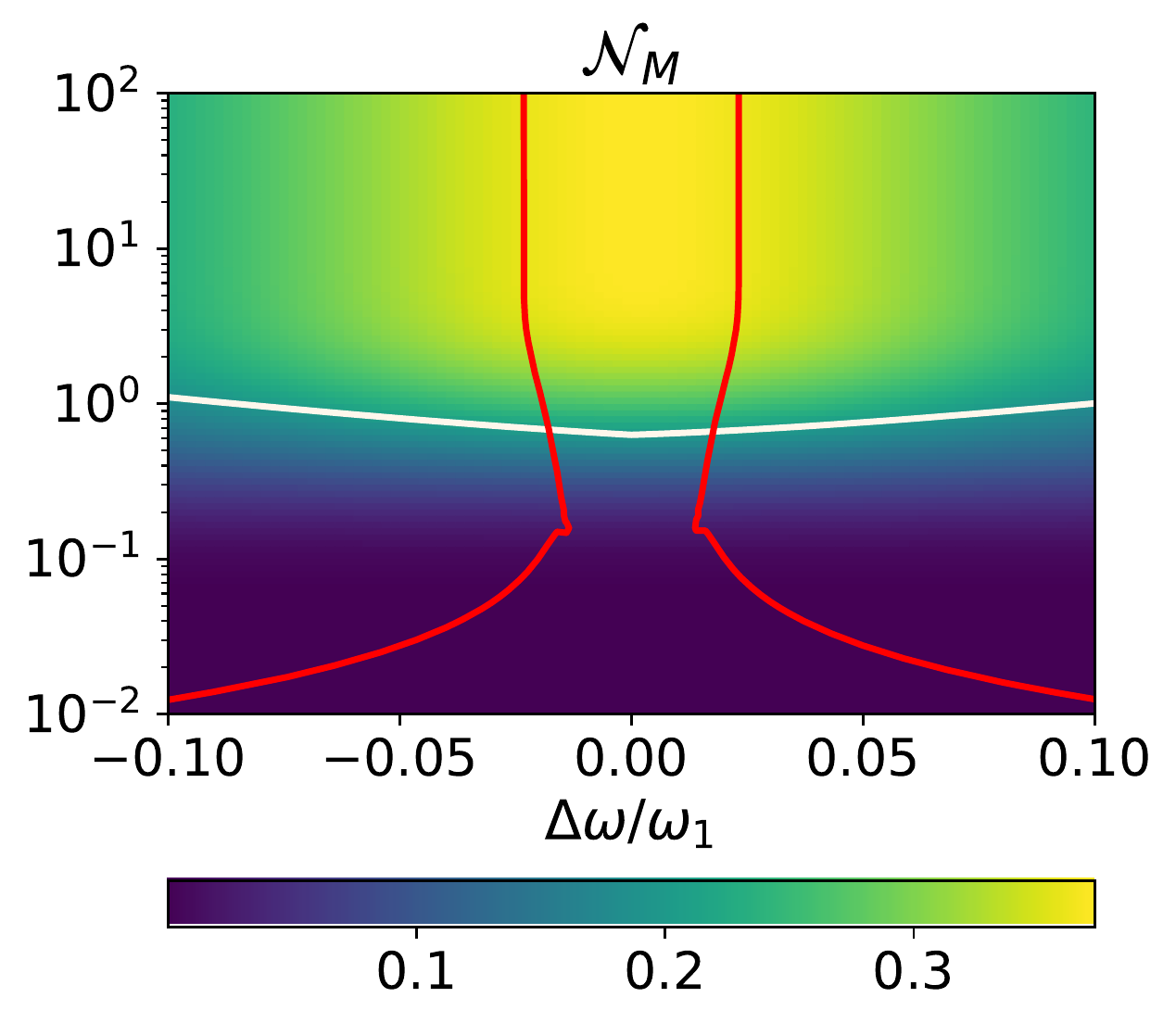}%
}
\subfloat[]{%
  \includegraphics[scale=0.34]{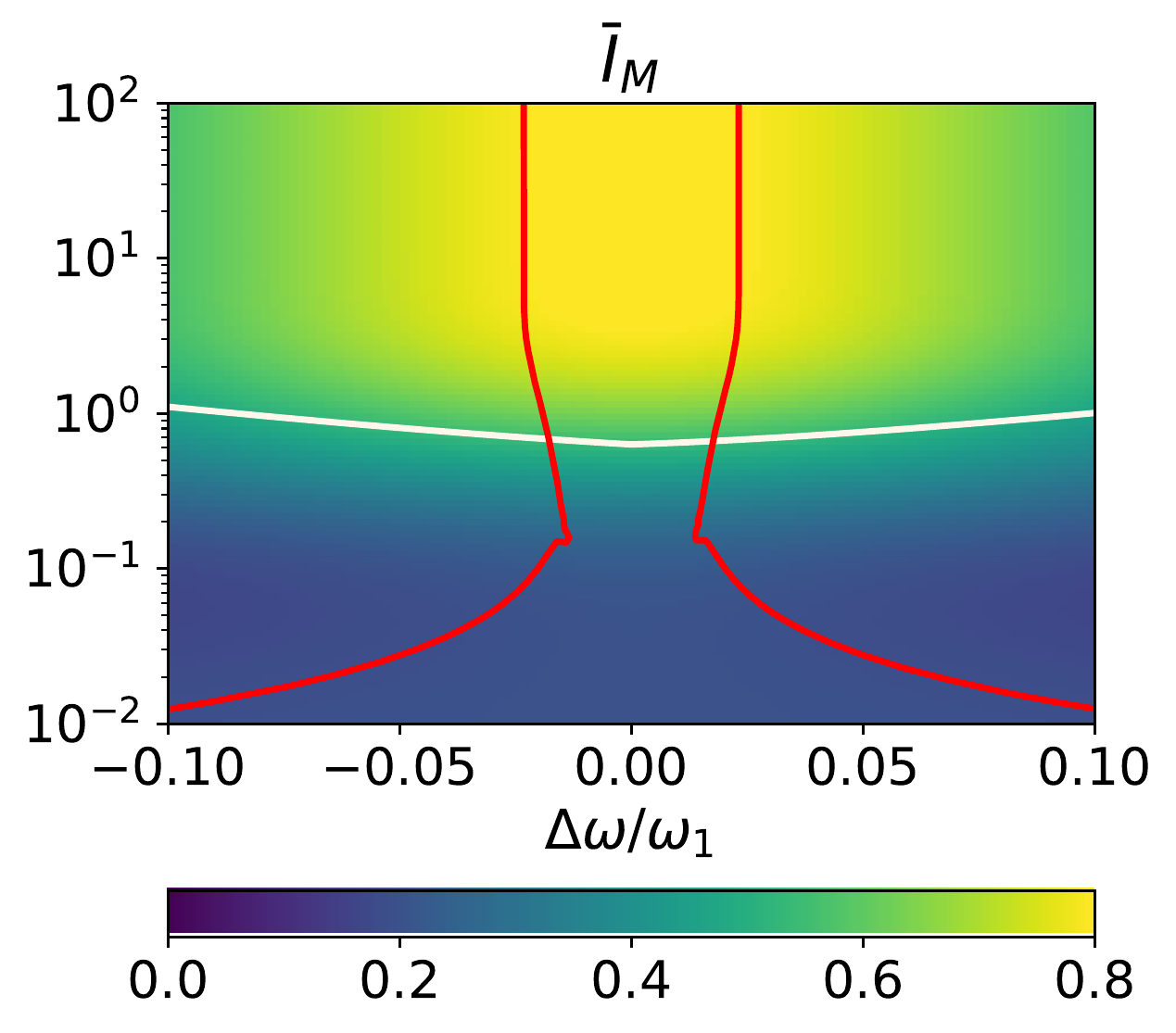}%
}
\caption{Measures of quantum synchronization, subradiance, entanglement generation and collectiveness in the balanced scenario ($g_1=g_2$), varying the detuning $\Delta\omega=\omega_1-\omega_2$ and the coupling constant $\mu$ (Figs.~(a)-(d)) or the inverse temperature $\beta$ (Figs.~(e)-(h)). Having fixed the quantity on the $y-$axis, the remaining parameters are set as $g_1=g_2=1$, $\mu=10^{-1.5}\hbar\omega_1$, $\beta=10/\hbar\omega_1$ and $T_1=3\times 10^5/\omega_1$. We have chosen an Ohmic spectral density of the environment (see Eq.~\eqref{eqn:ohmic}). The white line delimits the area in which $Syn$ is bigger than $1$, while the red line delimits the area in which $Sub$ is bigger than $1$ in Figs.~(c) and (d) and bigger than $100$ in Figs.~(g) and (h).}
\label{fig:detVSmuBeta}
\end{figure*}

\subsection{Numerical comparison in different scenarios}
\label{sec:comparison}

\begin{figure*}
\centering
\subfloat[]{%
  \includegraphics[scale=0.34]{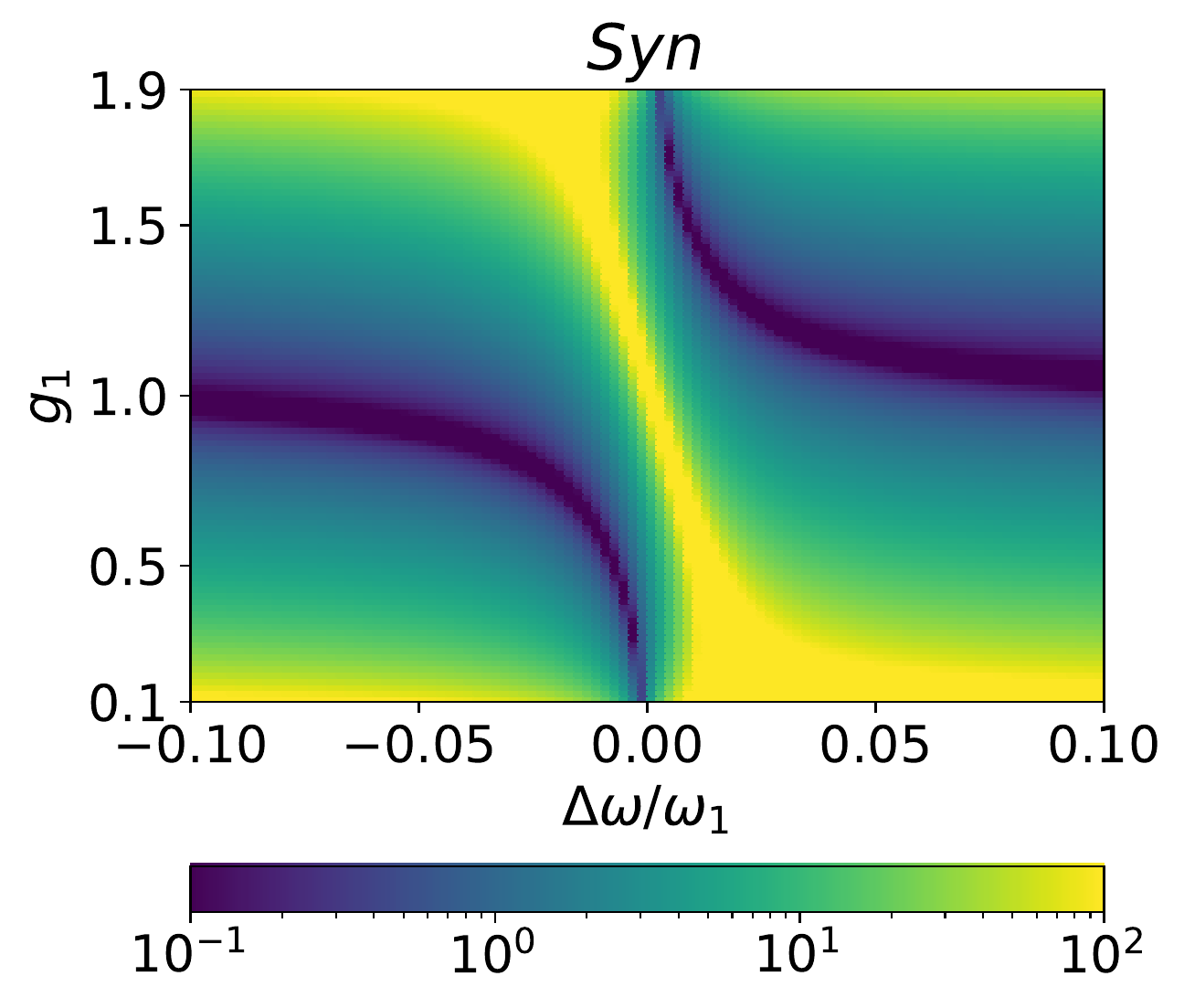}%
}
\subfloat[]{%
  \includegraphics[scale=0.34]{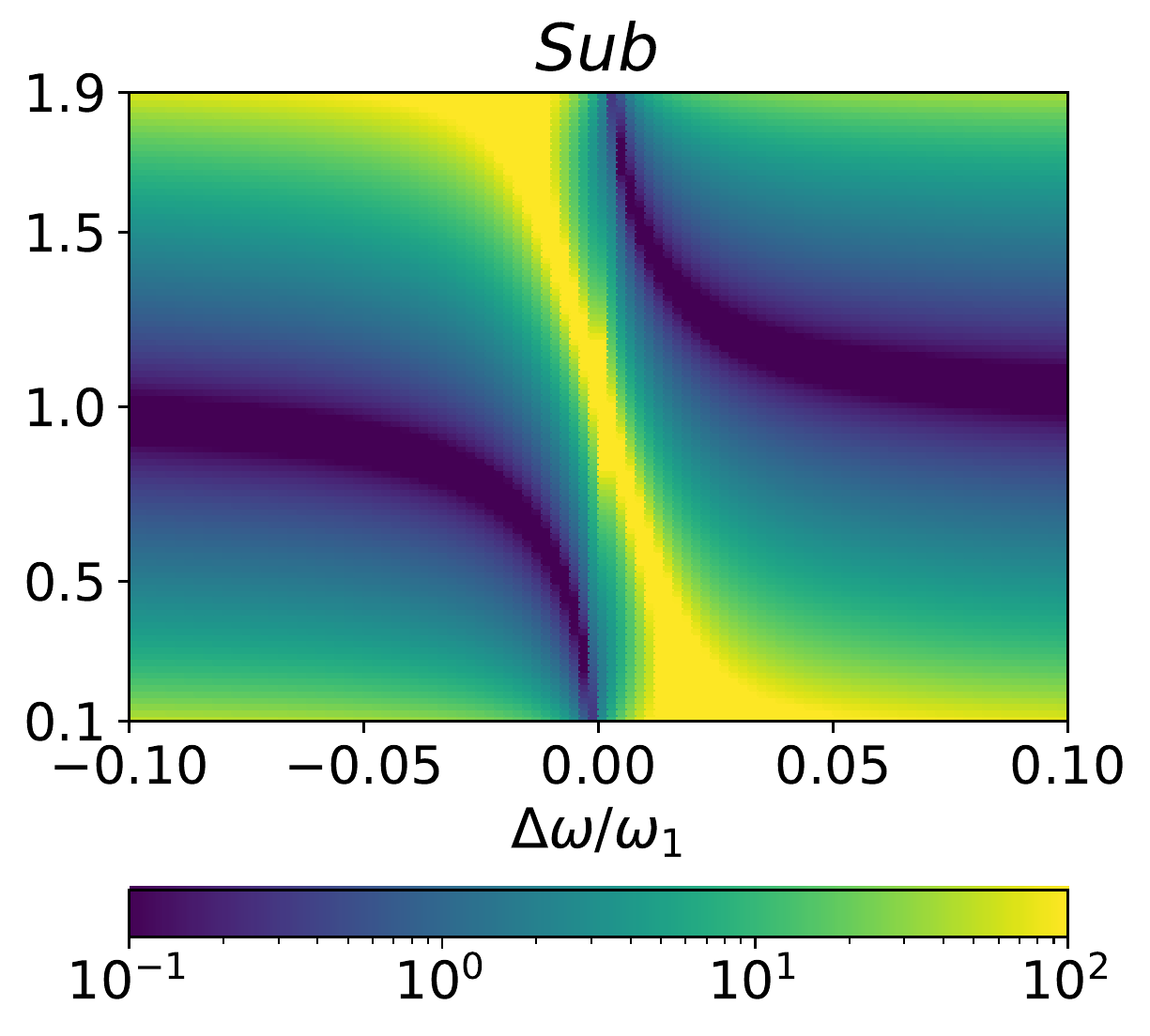}%
}
\subfloat[]{%
  \includegraphics[scale=0.34]{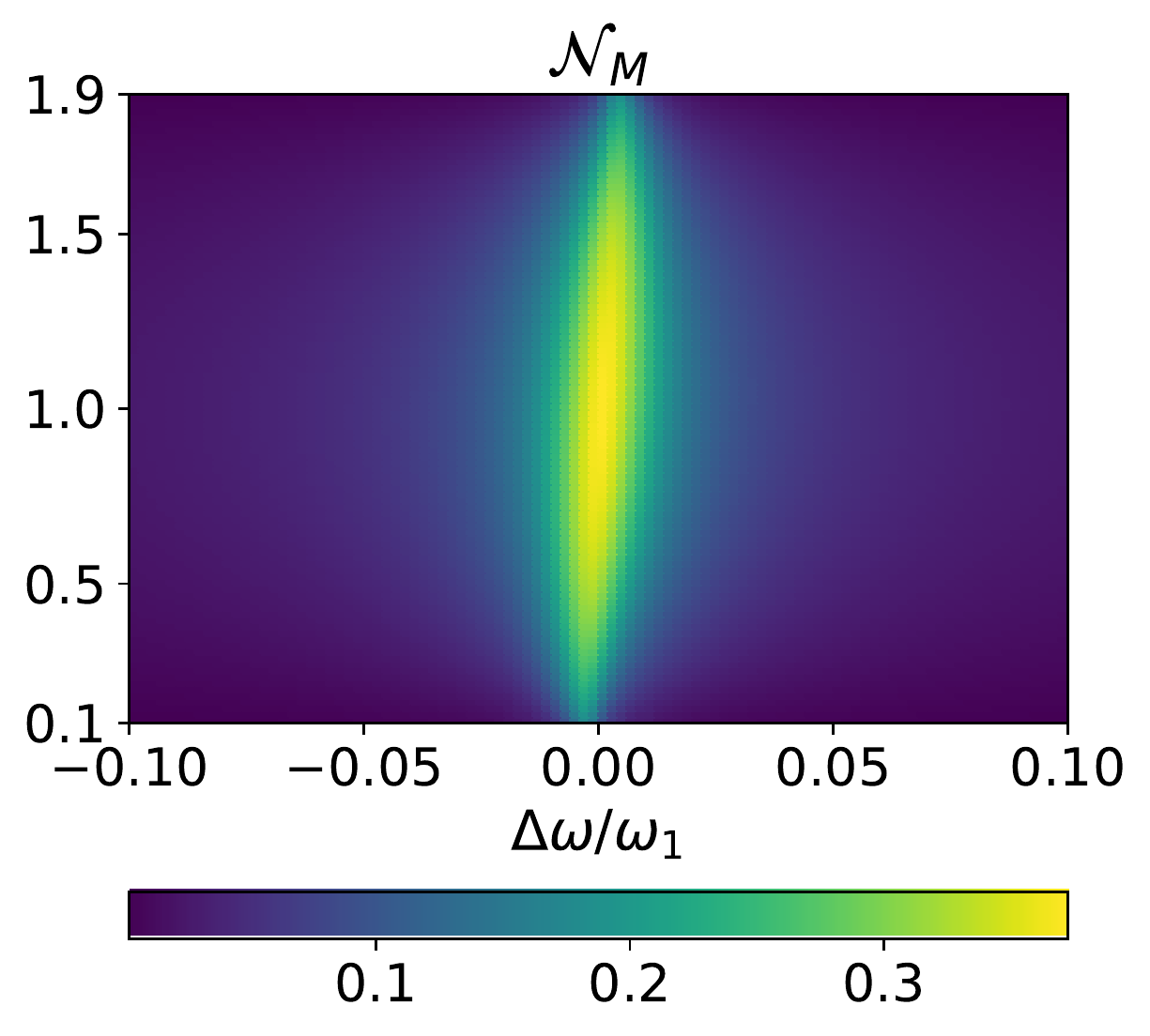}%
}
\subfloat[]{%
  \includegraphics[scale=0.34]{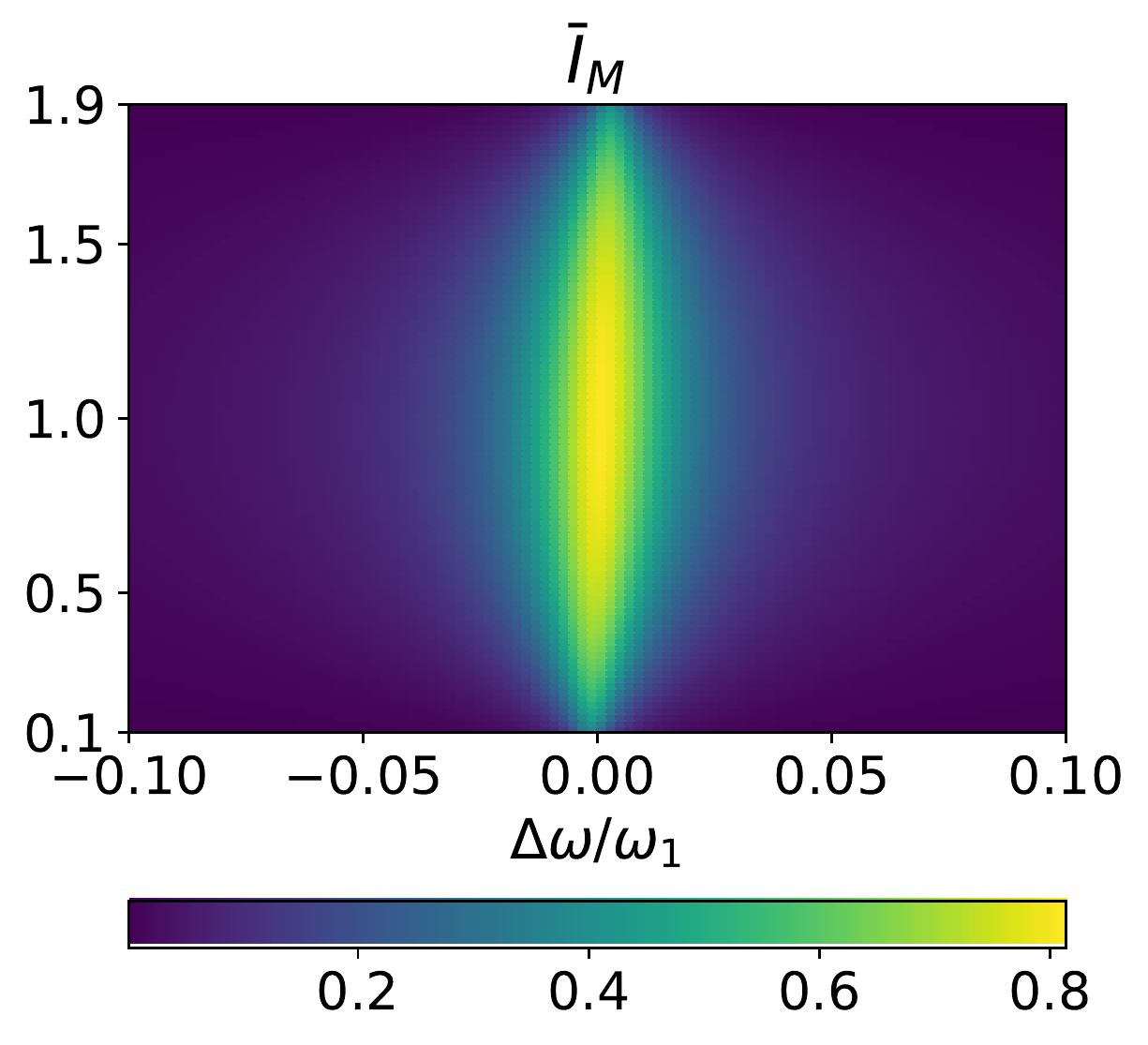}%
}
\caption{Measures of quantum synchronization, subradiance, entanglement generation and collectiveness in the presence of an unbalanced coupling of each qubit to the bath. We vary the detuning $\Delta\omega=\omega_1-\omega_2$ and the weight of the dissipative interaction $g_1$ (while $g_2=2-g_1$). The remaining parameters are set as $\mu=10^{-2}\hbar\omega_1$, $\beta=10/\hbar\omega_1$ and $T_1=3\times 10^5/\omega_1$. We have chosen an Ohmic spectral density of the environment (see Eq.~\eqref{eqn:ohmic}).} 
\label{fig:detVSg}
\end{figure*}

In the following, we will explore several situations by studying how the figures of merit vary for different parameters in the two qubits platform
introduced in Sec.~\ref{sec:model}.
We will analyze their behavior against the detuning, the bath temperature, the system-bath coupling, and the unbalancing of the coupling, also exploring the dependence on initial conditions. For the sake of clarity, we will first discuss the case $g_1=g_2$ and then move to 
the unbalanced case, corresponding to a situation in which collective dissipation acts with different strength on the two qubits. The results are displayed in Fig.~\ref{fig:detVSmuBeta} for the balanced case $g_1=g_2$, 
where the four indicators are plotted against the detuning (\ref{fig:detVSmuBeta}(a-d)) and against the temperature (\ref{fig:detVSmuBeta}(e-h)) 
and in Fig.~\ref{fig:detVSg} for the unbalanced case $g_1\neq g_2$. Further results are presented in Fig.~\ref{fig:further}. In all the scenarios, anticipating the relevant case for the platform we are going to introduce in Sec.~\ref{sec:experiment}, we set an Ohmic spectral density of the bath (see Eq.~\eqref{eqn:ohmic}).

\subsubsection{Balanced  couplings}

\begin{figure*}
\centering
\subfloat[]{%
  \includegraphics[scale=0.34]{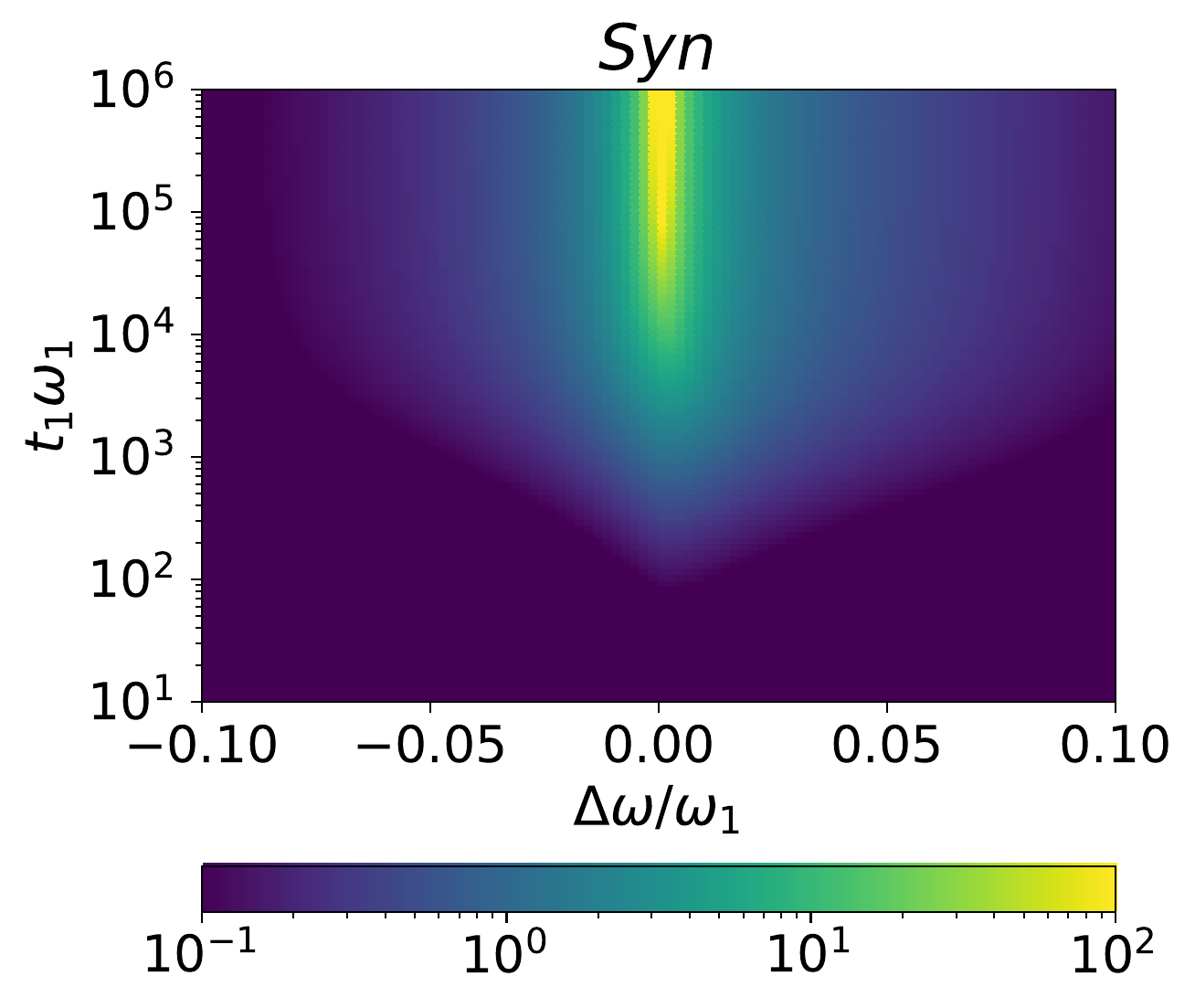}%
}
\subfloat[]{%
  \includegraphics[scale=0.34]{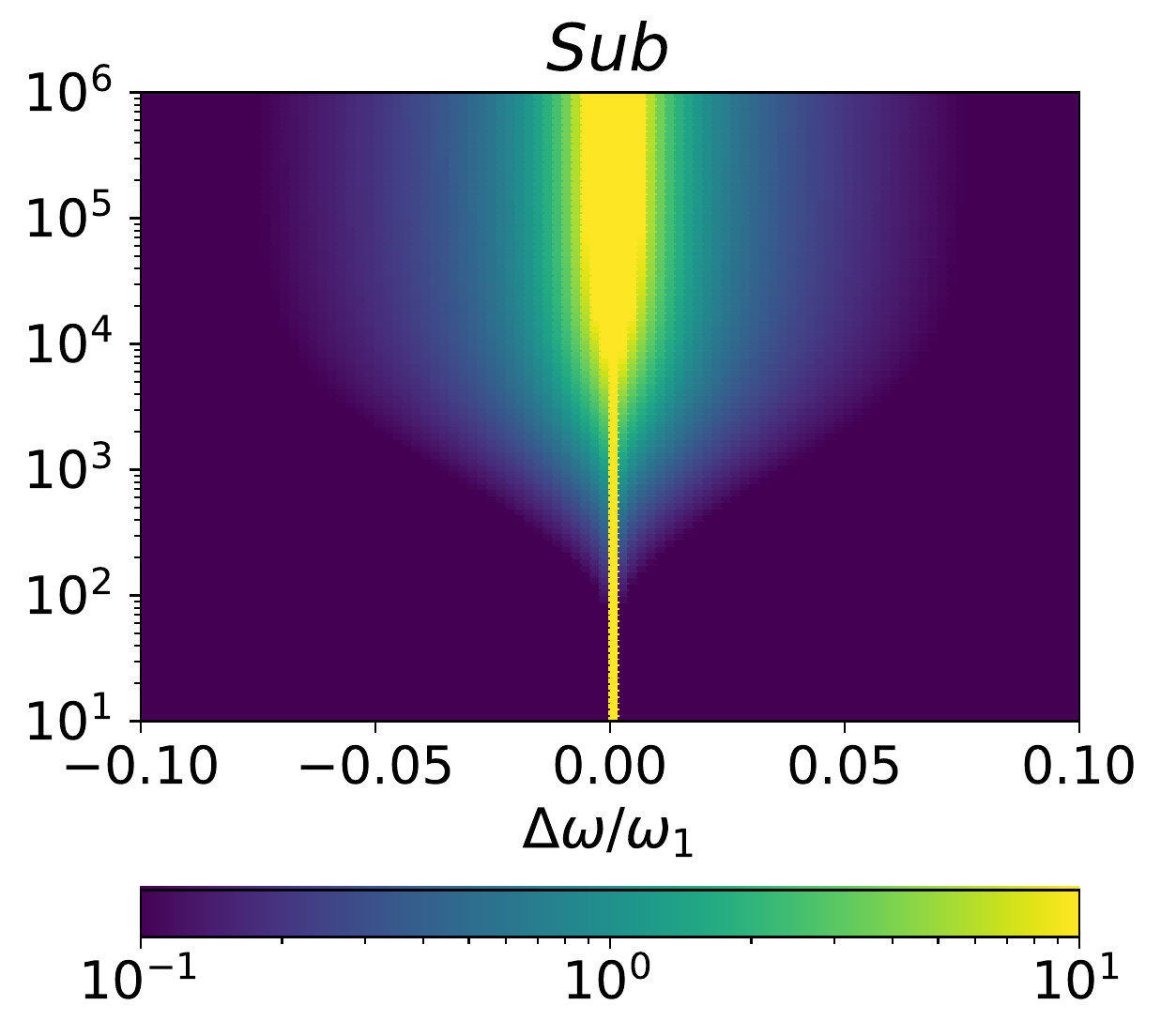}%
}
\subfloat[]{%
  \includegraphics[scale=0.34]{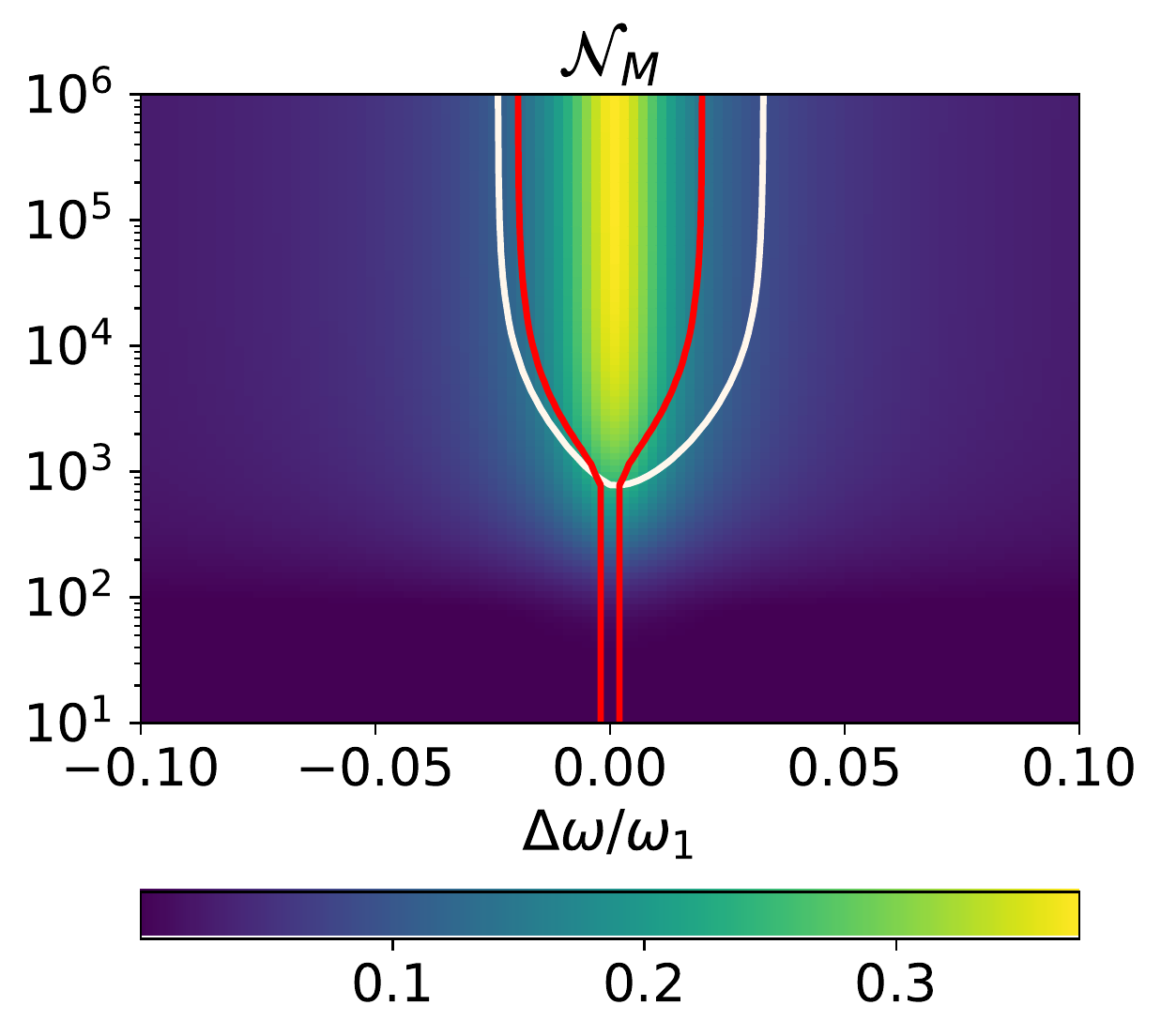}%
}
\subfloat[]{%
  \includegraphics[scale=0.34]{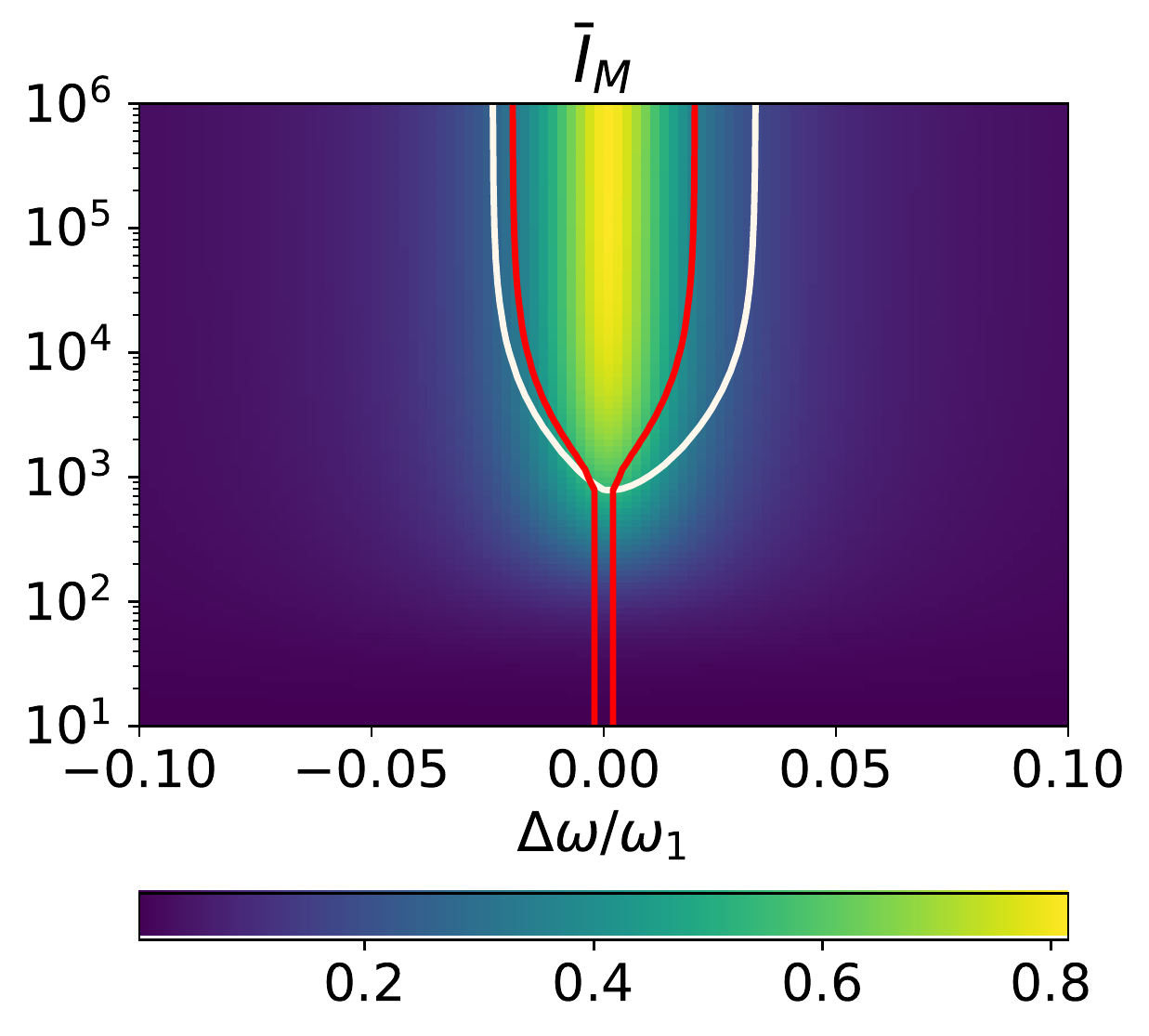}%
}\\
\subfloat[]{%
  \includegraphics[scale=0.34]{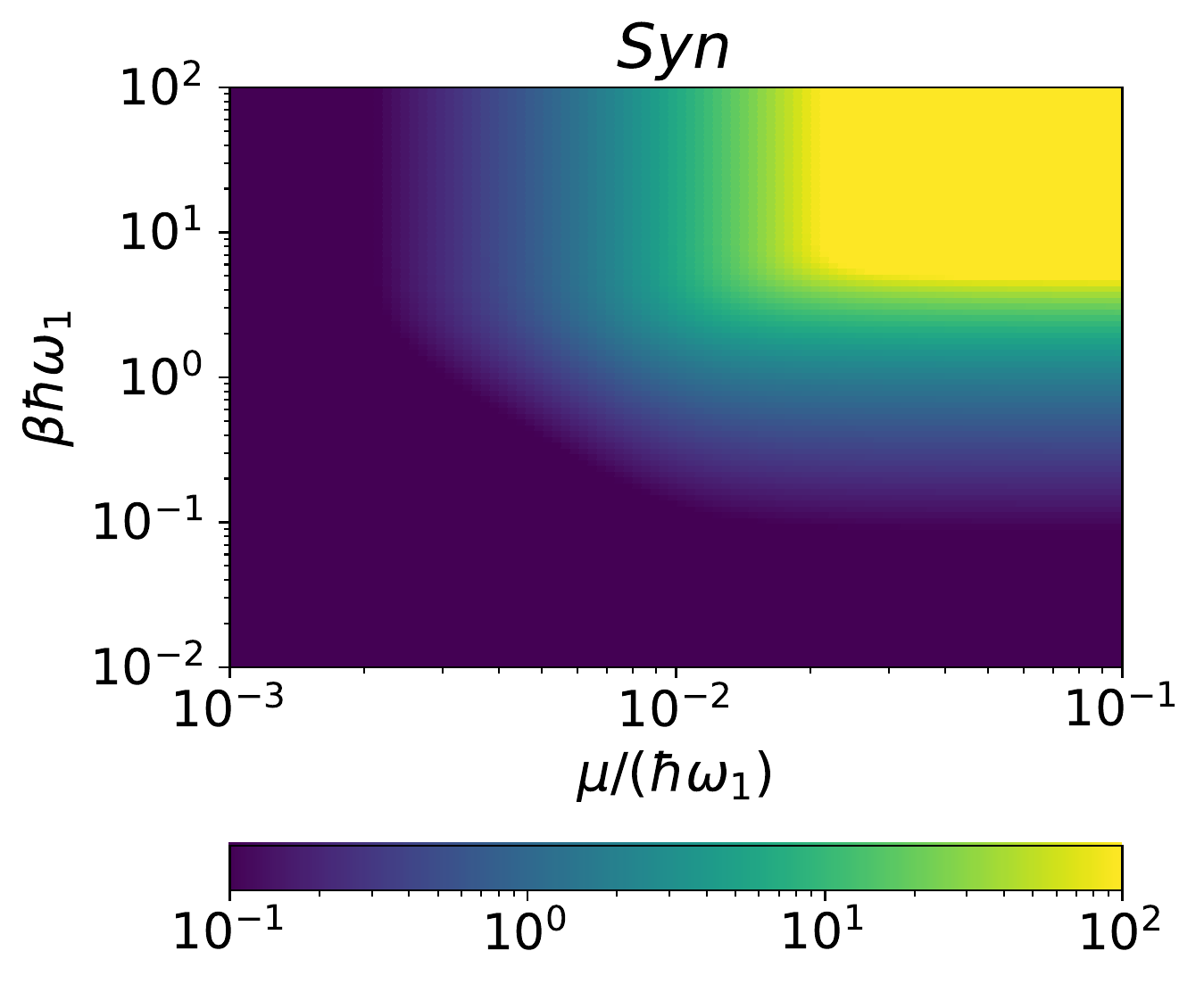}%
}
\subfloat[]{%
  \includegraphics[scale=0.34]{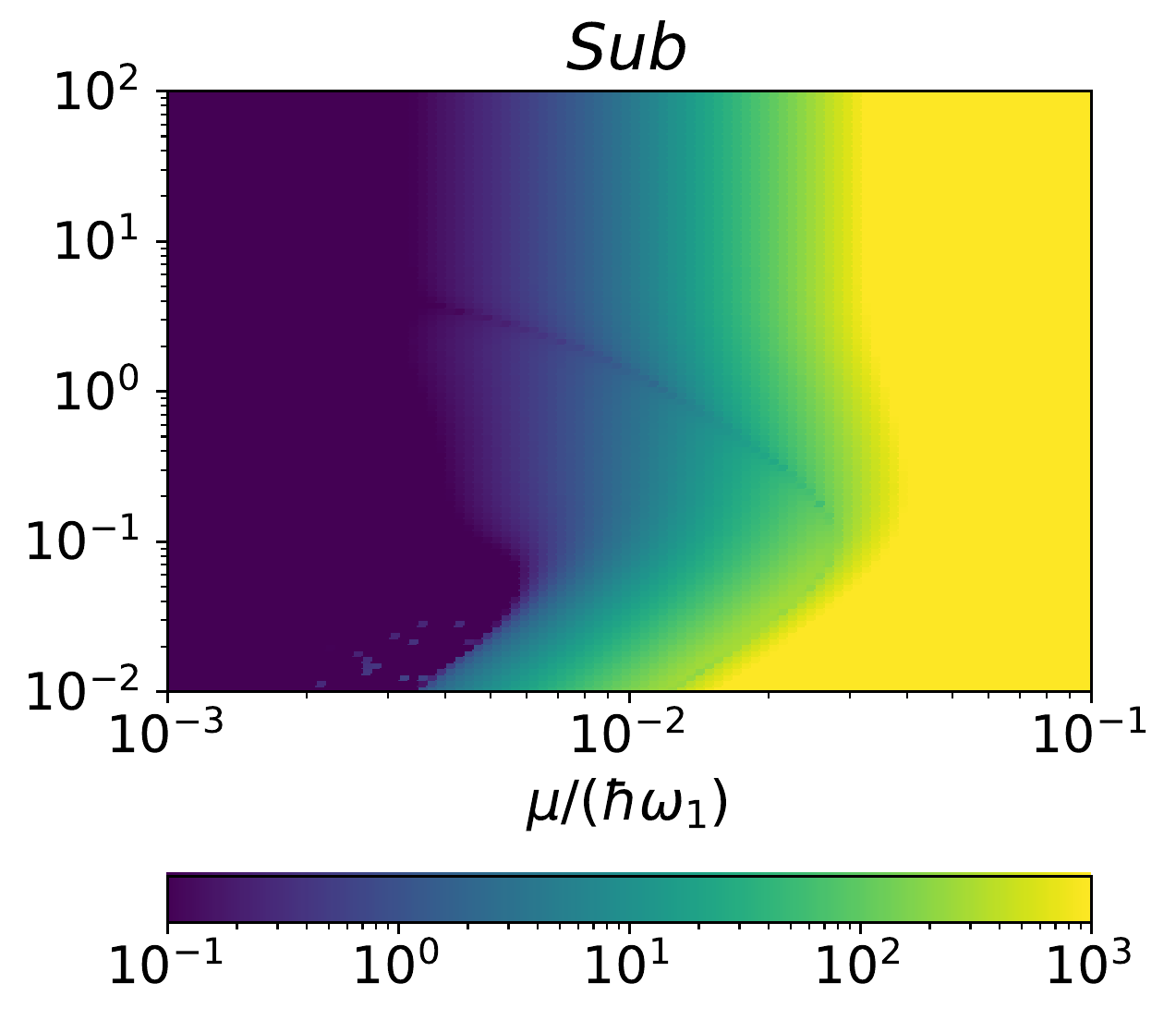}%
}
\subfloat[]{%
  \includegraphics[scale=0.34]{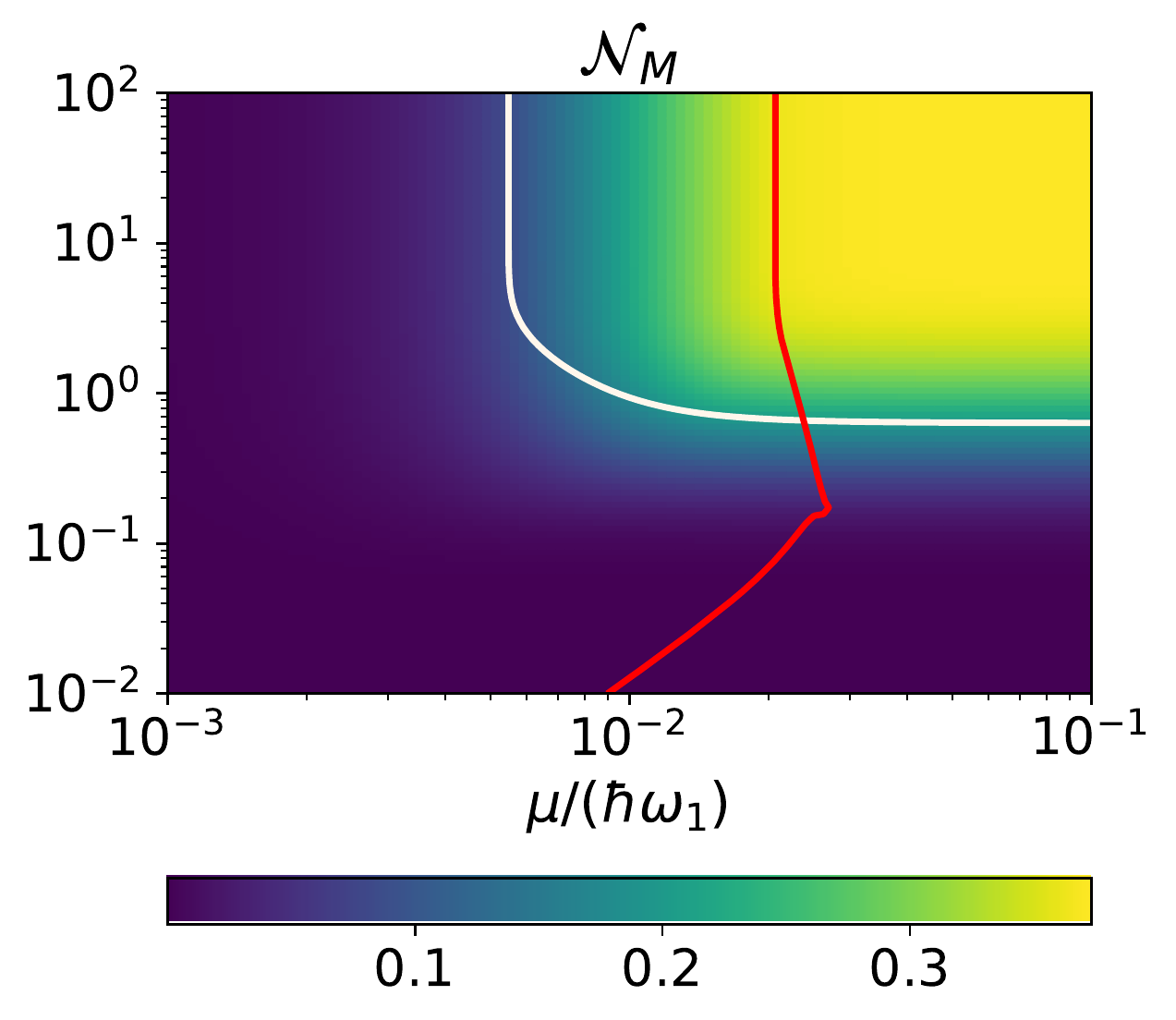}%
}
\subfloat[]{%
  \includegraphics[scale=0.34]{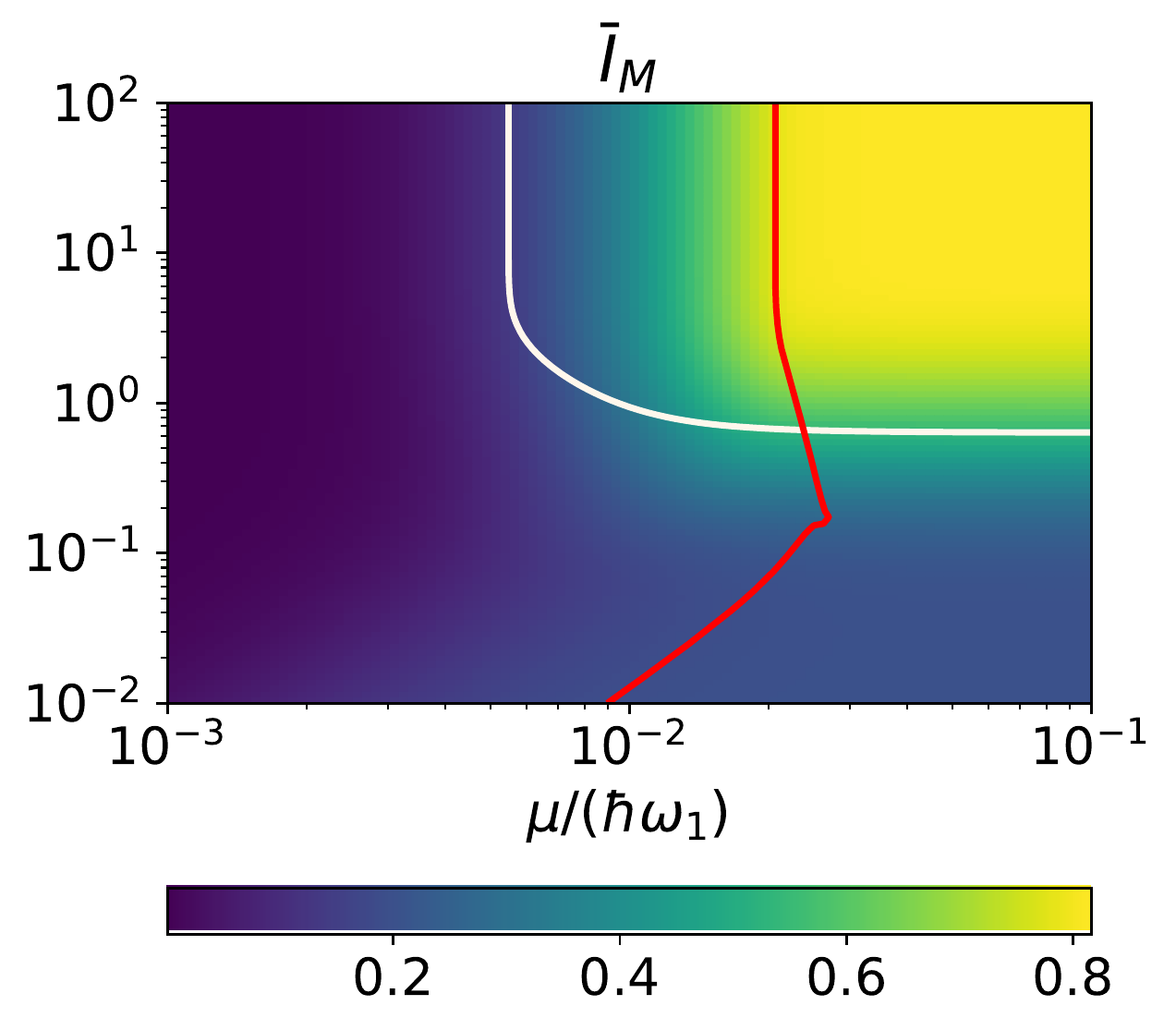}%
}\\
\subfloat[]{%
  \includegraphics[scale=0.34]{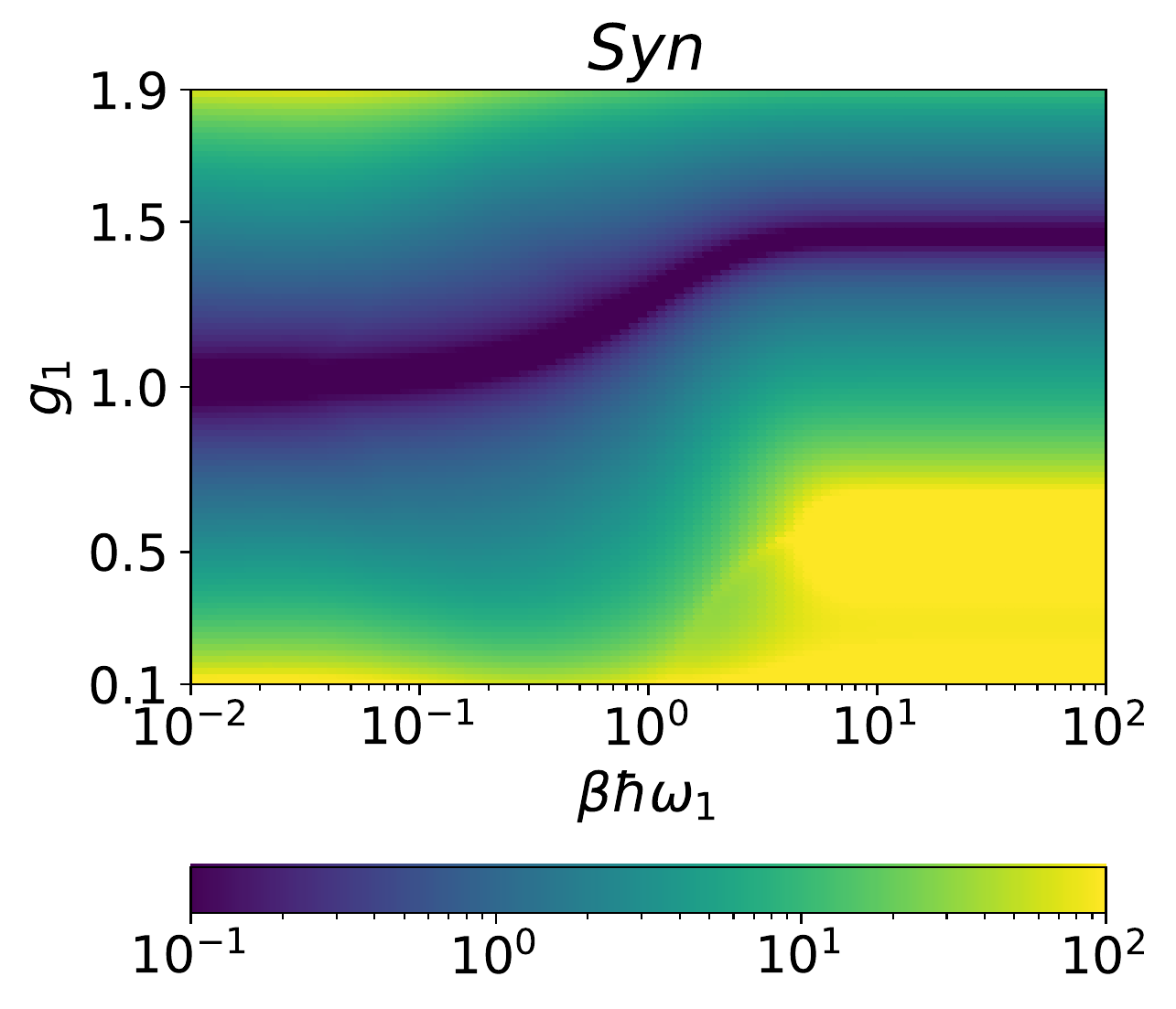}%
}
\subfloat[]{%
  \includegraphics[scale=0.34]{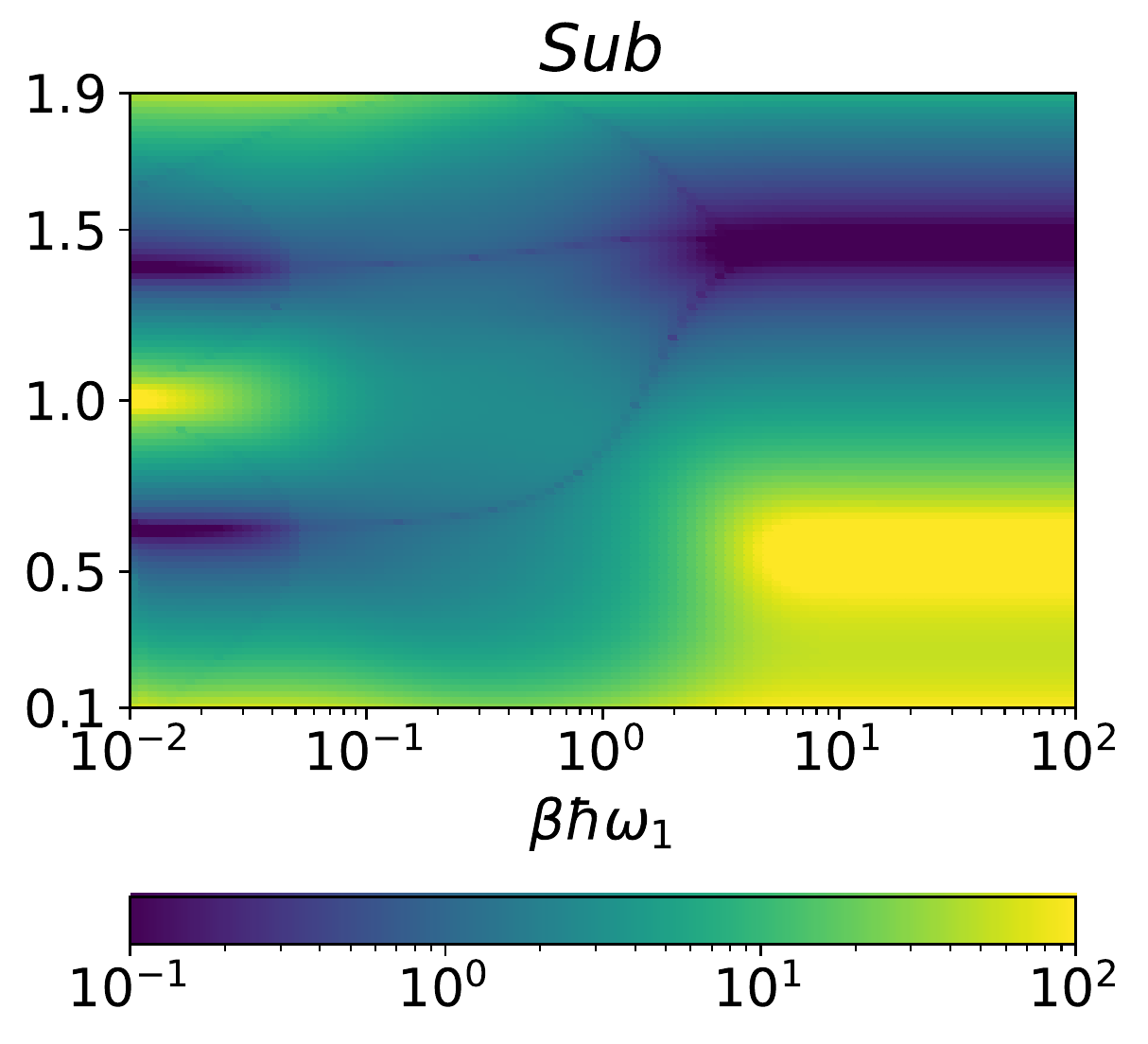}%
}
\subfloat[]{%
  \includegraphics[scale=0.34]{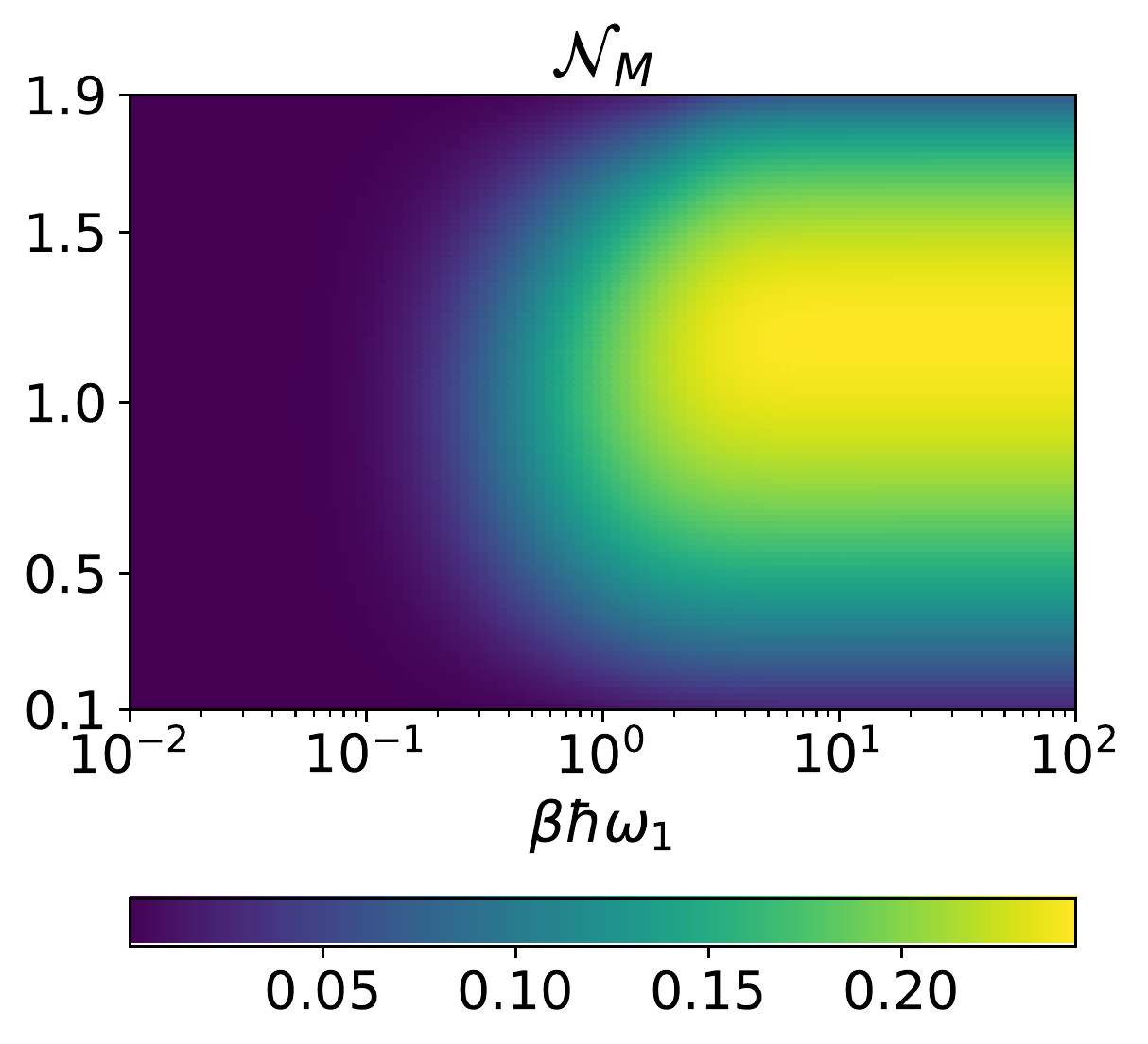}%
}
\subfloat[]{%
  \includegraphics[scale=0.34]{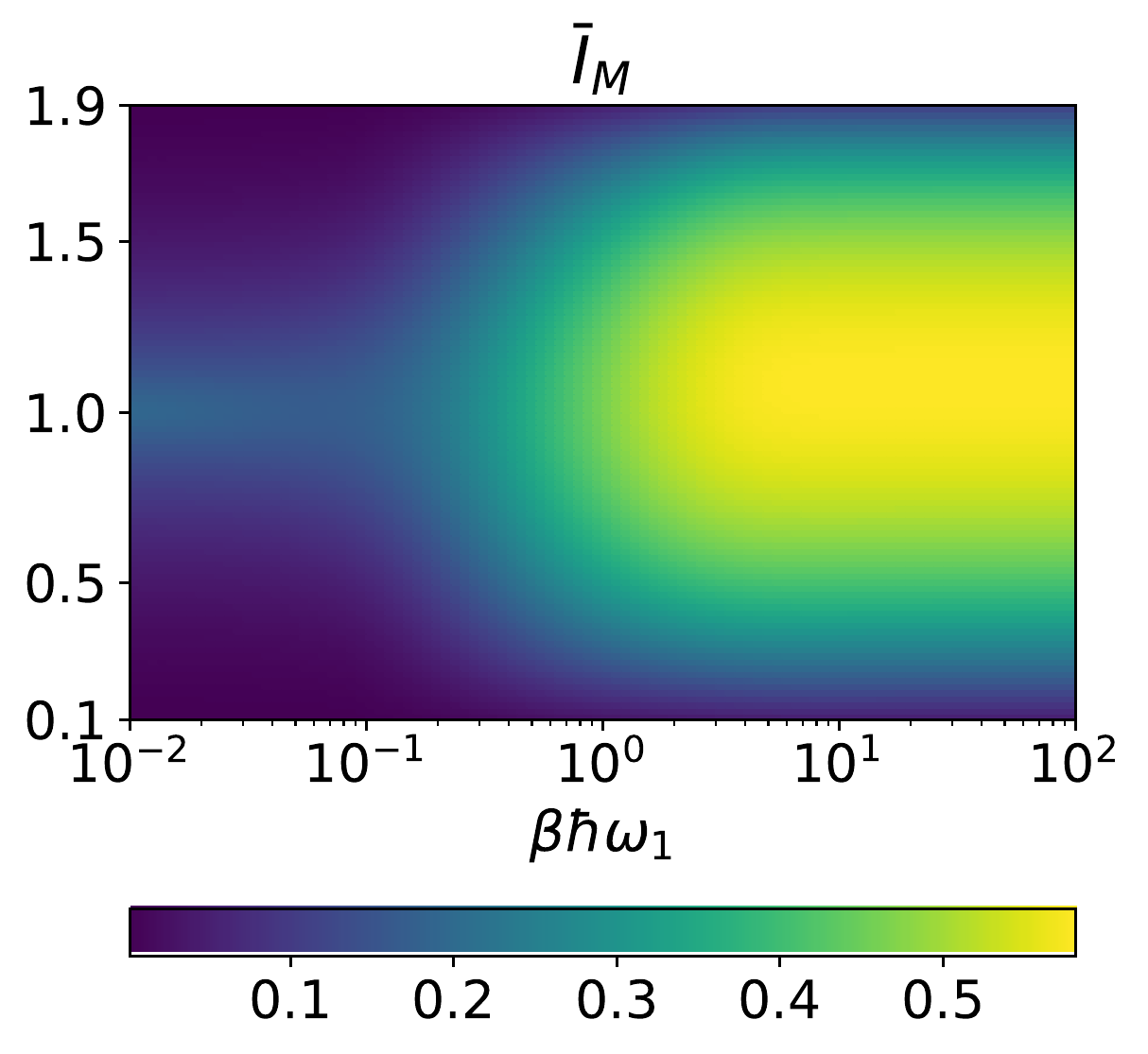}%
}
\caption{Measure of quantum synchronization, subradiance, entanglement generation and collectiveness in different ranges of parameters. In Figs.~(a)-(d) we vary the detuning $\Delta\omega=\omega_1-\omega_2$ and the local dissipative time $T_1$, having fixed $\mu=10^{-2}\hbar\omega_1$, $\beta=10/\hbar\omega_1$ and $g_1=g_2=1$. In Figs.~(e)-(h) we vary the coupling constant $\mu$ vs. the inverse temperature $\beta$, with $\Delta\omega=0.01\omega_1$, $T_1=3\times 10^5/\omega_1$ and $g_1=g_2=1$. In Figs.~(i)-(l) we study the unbalanced scenario, varying the weight $g_1$ (with $g_2=2-g_1$) and the inverse temperature $\beta$; the remaining parameters read $\mu=10^{-2}\hbar\omega_1$, $\Delta\omega=0.01\omega_1$ and $T_1=3\times 10^5/\omega_1$. We have chosen an Ohmic spectral density of the environment (see Eq.~\eqref{eqn:ohmic}). The white line in Figs.~(c)-(d) and~(g)-(h) indicates the parameter region in which $Syn>1$, while the red line delimits the area with $Sub>1$ in Figs.~(c)-(d) while $Sub>100$ in Figs.~(g)-(h).}
\label{fig:further}
\end{figure*}

Let us start analyzing the role played by the system-bath coupling: the higher 
the coupling strength, the stronger the effects of the common bath, and 
therefore the larger the four measures. This can be observed, for instance, in 
Figs.~\ref{fig:detVSmuBeta}(a)-(d) and in Figs.~\ref{fig:further}(e)-(h). If 
$\mu$ is too weak, the unitary evolution driven by the system Hamiltonian will 
play the only relevant role in the dynamics, and no collectiveness, 
entanglement, synchronization or subradiance will appear. For instance, with the 
parameters employed in Figs.~\ref{fig:detVSmuBeta}(a)-(d) (i.e. $g_1=g_2=1$, 
$\beta=10/\hbar\omega_1$, $T_1=3\times 10^5/\omega_1$), we observe that for 
$\mu\lessapprox 10^{-2.5}\hbar\omega_1$ and $\Delta\omega\approx 0.01\omega_1$ neither synchronization 
nor subradiance emerge before thermalization, while both negativity and 
collectiveness are negligible.

As for the detuning, in all scenarios, $\mathcal{N}_M$ and $\bar{I}_M$ decrease
as soon as $\Delta \omega$ increases, consistently with a vanishing dissipative coupling.
Indeed, in spite of the presence of a common bath, in this regime the dynamics
can be well approximated by a full secular master equation
dominated by local instead of collective dissipation \cite{Cattaneo2019}.  
This also occurs
when varying the value of the local relaxation time $T_1$ 
(Fig.~\ref{fig:further}(c)-(d)).
 As shown in Figs.\ref{fig:detVSmuBeta}(a)-(b)-(e)-(f), synchronization and subradiance
display the same behavior: in Figs.~\ref{fig:detVSmuBeta}(c)-(d) we have drawn a 
line demarcating the regions of parameters in which $Syn$ and $Sub$ are bigger 
than $1$, and the same in Figs.~\ref{fig:detVSmuBeta}(g)-(h) with the difference 
that the red line indicates a value of $Sub>100$. In 
Figs.~\ref{fig:detVSmuBeta}(c)-(d) (at fixed inverse temperature 
$\beta=10/\hbar\omega_1$) we see that these regions coincide with the ones where $\mathcal{N}_M$ and $\bar{I}_M$ display stronger values. We observe that
synchronization shows a slight asymmetry between the scenarios with $\Delta\omega>0$ 
and $\Delta\omega<0$. This is due to the fact that $Syn$ is particularly 
sensible to the change of the absolute value of the frequencies $\omega_1$ and 
$\omega_2$, and not only of the detuning $\Delta\omega$.
 
More complex is the dependence with the temperature, as depicted in Figs.~\ref{fig:detVSmuBeta}(e)-(h) and 
Figs.~\ref{fig:further}(e)-(h). In this case, 
the measure of subradiance displays notable differences with respect to the 
other ones: it is greatly enhanced by higher temperatures, as found analytically, unlike the other phenomena.  
We indeed observe that, even for the case with finite detuning, the slowest decay rate is not significantly 
affected by the value of the temperature, while all the remaining ones are speeded up when temperature increases, and therefore $Sub$ is enhanced for higher 
temperatures. 
As for synchronization, in agreement with the analytical limit and what was  partially 
discussed in the literature 
\cite{Cabot2019,Karpat2020}, we observe that while a higher temperature speeds 
up the relevant decay rates  (formally the eigenvalues of the block 
$\mathcal{L}_1$, see Eq.~\eqref{eqn:sigmaX}), it does not increase the gap 
between the two slowest eigenvalues. As a consequence, the synchronization time 
does not change considerably, while the relaxation time occurs way before than 
at lower temperatures. The lack of a significant transient where synchronization can be observed is indeed captured by a decrease of the measure of  Eq.~\eqref{eqn:syncMeasure}. Furthermore, entanglement smoothly disappears as temperature increases, vanishing for $\beta\hbar\omega_1\ll 1$, in agreement with the analytical limit of Sec.~\ref{sec:anL}. Finally, we observe that the collectiveness decreases as well 
when the temperature increases, although, contrary to the negativity and 
synchronization, its value never vanishes for $T\rightarrow\infty$. Indeed,  
even at infinite temperature the master equation describing the dynamics driven by a common bath is different from the one driven by two independent local 
baths.

\subsubsection{Unbalanced  couplings}
{The maximal collective effect is achieved when the coupling between the qubits and the bath is balanced, i.e. $g_1=g_2$. A less relevant scenario appears when this is not true anymore, and the collectiveness of the evolution inevitably decreases. Let us analyze this case:} the bigger the imbalance between the weights of the dissipative coupling  $g_1$ and $g_2$, the smaller the collective effects in the dynamics. Indeed, both collectiveness and negativity decrease as the imbalance increases (see Figs.~\ref{fig:further}(k)-(l) and~\ref{fig:detVSg}(c)-(d)). If  we weaken the coupling of one of the two qubits by varying $g_1$ and $g_2$, we monotonically decrease the values of $\mathcal{N}_M$ and $\bar{I}_M$, reaching zero in the trivial limit $g_1\rightarrow 0$, $g_2\rightarrow 2$.
{On the other hand, a significative imbalance between the weights of the dissipative interaction $g_1$ and $g_2$ breaks the possibility of using synchronization and subradiance as signatures of entanglement generation. Indeed,} the appearance of  synchronization and subradiance depends not only on the collectiveness of the evolution, but also on the gap between the eigenvalue of the slowest eigenmode and all the remaining ones.  By unbalancing the coupling, we may enhance this gap, and therefore increase the values of $Syn$ and $Sub$ (as for instance already discussed in Ref.~\cite{Giorgi2016}). This behavior is observed in Figs.~\ref{fig:detVSg} and~\ref{fig:further}(i)-(j), which shows that, by unbalancing $g_1$ and $g_2$, the separation between the eigenvalues of both synchronization and subradiance is enhanced in a stronger way than the collectiveness of the relevant eigenmodes is decreased,  and therefore we obtain higher values of $Syn$ and $Sub$. 
Finally, let us comment that, in the unbalanced scenario, the behavior of synchronization and subradiance when varying the temperature is now completely different from the balanced case (see Figs.~\ref{fig:further}(i)-(j)). Indeed, the arguments we used in the previous section to understand the behavior of the figures of merit as a function of the temperature are now not valid anymore, and, for example, for some values of $g_1$ and $g_2$ we obtain a higher synchronization for higher temperatures.

\subsubsection{Dependence on the initial conditions}
\label{sec:initCond}
Let us finally try to understand whether  our results depend on the specific initial states we have chosen. In order to do so, for the case of synchronization and subradiance, we have computed a modified form of their figure of merit in which we do not consider the initial conditions. In particular, we have set each $p_{0j}^{(1)}$ and $p_{0j}^{(0)}$ equal to $1$ in Eqs.~\eqref{eqn:syncTime} and~\eqref{eqn:subTime}, and we have computed these slightly modified measures in all the scenario discussed before. We have found that the results without taking into account the initial conditions mimic the ones depicted in Figs.~\ref{fig:detVSmuBeta},~\ref{fig:detVSg} and~\ref{fig:further}, with no exceptions. 

Focusing on the entanglement generation, we can explore different initial conditions by considering the states discussed in Appendix~\ref{sec:entProof}. We have found that the sufficient conditions for having entanglement at time $t\rightarrow 0^+$ starting from a given initial state, actually can be associated to the behavior of the entanglement generation throughout the whole evolution. For instance, we observe that no entanglement is generated by the bath when starting the dynamics in the pure states $\ket{ee}$ or $\ket{gg}$, which as proven in Appendix~\ref{sec:entProof} can never display entanglement at $t\rightarrow 0^+$. We have then considered the initial states $\ket{eg}$ and $\ket{ge}$. We observe that, in both cases, the maximum value of negativity generated during the evolution is higher than the one observed in Figs.~\ref{fig:detVSmuBeta},~\ref{fig:detVSg} and~\ref{fig:further}, where the choice was $\rho_S(0)=\rho_{C}$. Curiously, the sufficient conditions of Appendix~\ref{sec:entProof} also suggest that these states have, in general, a stronger power of generating entanglement at $t\rightarrow 0^+$. Moreover, we observe that the asymmetry between qubit $1$ and qubit $2$ is reflected in the behavior of $\mathcal{N}_M$ as a function of the detuning: for instance, if we start the dynamics in $\ket{eg}$ we find stronger entanglement for $\Delta\omega=+x$ than for $\Delta\omega=-x$, and viceversa. Still, the dependence as a function of $\Delta\omega$ does not change, and all the results of the previous discussion hold. In all the other scenarios the behavior of $\mathcal{N}_M$ is analogous to the one depicted in the figures of the main text. We therefore conclude that the dependence on the initial conditions does not affect the findings of our work, and the same conclusions can be drawn for general scenarios.

\subsection{Extensions to multi-qubit systems}
In Sec.~\ref{sec:anL} we have discussed how the four figures of merit behave in some particular limits, which are valid independently of the chosen model or physical scenario, and hold for any system of two qubits dissipatively coupled to a thermal bath. In Sec.~\ref{sec:comparison} we have performed an extensive numerical investigation for different scenario inspired by two transmon qubits immersed in an Ohmic thermal bath. The extensions of these results to the multi-qubit case  would deserve a separate study, but we can anticipate some issues in this more complex scenario. Adding a single qubit, one may take inspiration from the discussion in Ref.~\cite{Manzano2013a}, focused on three harmonic oscillators. Here, the different strengths of each coupling interaction with the bath play a relevant role, and different conditions for decoherence-free subspaces may appear in some parameter manyfolds. Furthermore, it is crucial to distinguish between synchronization of only two of the three qubits immersed in a common bath and synchronization of the whole system. Correspondingly, one may decide to compare it with multipartite entanglement \cite{Horodecki2009} of the system, or just bipartite entanglement in a qubit pair (similar issues have been investigated in Ref.~\cite{Valido2015}). A recent work also studies different synchronization features in a three-qubit system \cite{Karpat2020}. The case of multiple qubits is even more complex, and one may also address the coexistence of common and local baths acting on a subsystem of the qubits only, as studied for harmonic oscillators in Refs.~\cite{Manzano2013,Cabot2018}.

\section{An experimental proposal to test our findings}
\label{sec:experiment}

\begin{figure}
\centering
\includegraphics[scale=0.125]{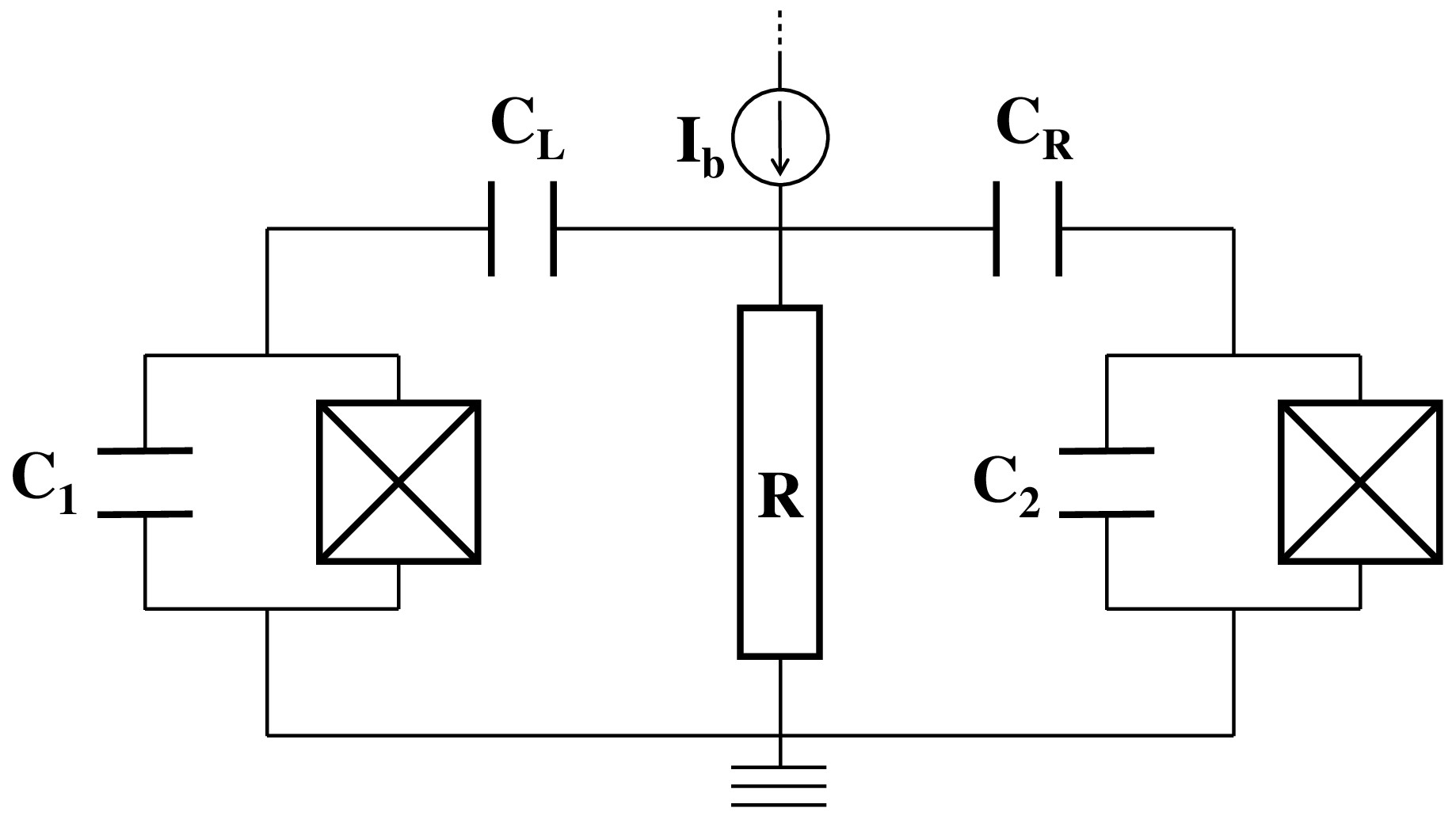}
\caption{Circuit diagram representing two transmon qubits, characterized by the capacitances $C_1$ and $C_2$, capacitively coupled to the same resistor $R$, respectively through the capacitors $C_L$ and $C_R$. The resistor plays the role of a common bath with Ohmic spectral density, that can be heated up by an external bias current $I_b$.}
\label{fig:circuit}
\end{figure}

We propose here an experimental platform that would allow one to test our predictions in a controllable environment and beyond mathematical approximations, 
eventually also allowing for the exploration of further regimes. The phenomena we have discussed can be observed by using two superconducting qubits, in a simple experiment as depicted in Fig.~\ref{fig:circuit}: two transmon qubits of frequency $\omega_1$ and $\omega_2$ are coupled to the same resistor through the capacitors $C_L$ and $C_R$. Let us assume that  the qubits have total capacitances (shunt plus intrinsic junction capacitance) $C_1$ and respectively $C_2$. In the weak coupling limit $C_L,C_R \ll C_1,C_2$, using standard quantum network analysis in the presence of dissipation \cite{Parra-Rodriguez2018}, we obtain 
the Hamiltonian Eq.~\eqref{eqn:HamiltonianTotal}. Here the resistor plays the role of a thermal bath with an Ohmic spectral density proportional to the resistance $R$. The parameters $\mu$, $g_1$ and $g_2$ can be tuned by varying the values of the capacitors in the circuit. Indeed, we obtain: 
\begin{equation}
\label{eqn:g1g2}
\begin{split}
&g_1=\frac{C_L(C_1+C_2)}{(C_L+C_1)(C_L+C_R)},\\
&g_2=\frac{C_R(C_1+C_2)}{(C_R+C_2)(C_L+C_R)},
\end{split}
\end{equation}
and $\mu$ is a constant with the units of energy, proportional to the resistance and to $(C_L+C_R)/(C_1+C_2)$, that with the above assumption satisfies the weak coupling limit. The temperature of the thermal bath is the effective temperature at which the resistor is dissipating energy, typically some tens of millikelvin.

\begin{figure*}
\centering
\includegraphics[scale=0.16]{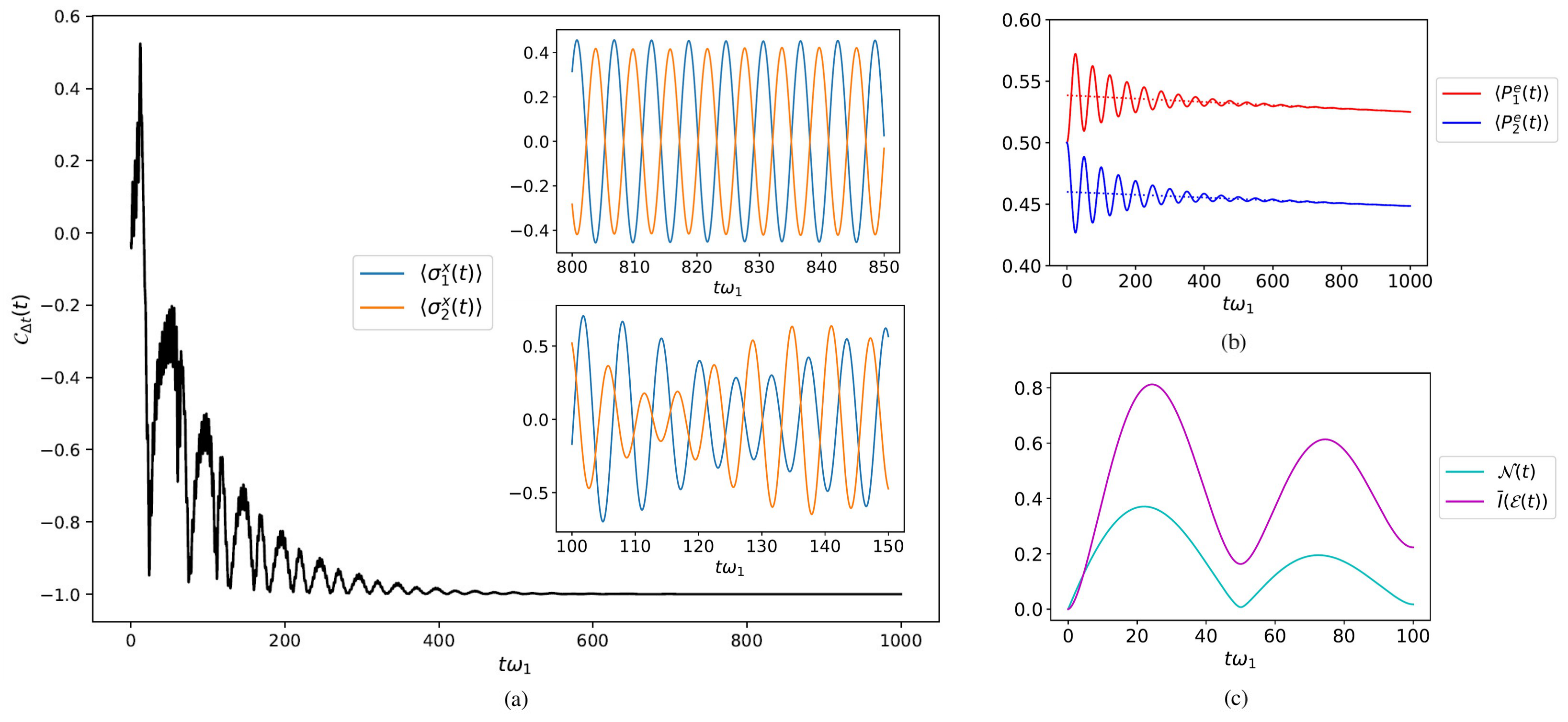}%
\caption{Evolution of different figures of merit in the scenario discussed in Sec.~\ref{sec:experiment}. (a): Pearson coefficient as a function of time, with $\Delta t=7/\omega_1\approx 0.2\,ns$. The insets depict the evolution of $\langle\sigma_1^x(t)\rangle$ and $\langle\sigma_2^x(t)\rangle$ in different time intervals. {(b): Dynamics of $\langle P_1^e(t)\rangle$ (solid red) and $p_{06}^{(0)}h_{61}^{(0)}+p_{05}^{(0)}h_{51}^{(0)}e^{\lambda_5^{(0)}t}$ (dotted red) for qubit 1, and $\langle P_2^e(t)\rangle$ (solid blue) and $p_{06}^{(0)}h_{62}^{(0)}+p_{05}^{(0)}h_{52}^{(0)}e^{\lambda_5^{(0)}t}$ (dotted blue) for qubit 2. The coefficients $p_{0j}^{(0)}$ and $h_{jk}^{(0)}$ are defined in Eq.~\eqref{eqn:sigmaZ}, and their combinations (dotted lines) represent the decays of the subradiant mode of each qubit as a function of time.}
(c): Evolution of negativity $\mathcal{N}(t)$ and collectiveness $\bar{I}(\mathcal{E}(t))$.}
\label{fig:evolution}
\end{figure*}

Let us consider a specific set of parameters: according to the discussion in the previous section, synchronization and subradiance are good signatures of entanglement generation in the balanced case, i.e. $g_1=g_2$. To obtain so, we choose $C_L$ and $C_R$ such that $C_L/(C_L+C_1)=C_R/(C_R+C_2)$. As reasonable qubit frequencies, we choose $\omega_1=2\pi\times 5$ GHz and $\omega_2=2\pi\times 4.95$ GHz, so $\Delta\omega=0.01\omega_1=2\pi\times 0.05$ GHz. In this configuration, state preparation can be achieved by standard methods in circuit quantum electrodynamics. Single qubit operations can be realized by applying microwaves tones in resonance with the corresponding qubit frequency. Since the coupling is fixed by the choice of the capacitors, two-qubit operations such as the CNOT needed to prepare the subradiant state can be realized naturally in this configuration by cross-resonance. In this method, two microwave pulses in resonance with the other qubit's frequency are applied simultaneously, which results in an effective tunable qubit-qubit coupling \cite{paraoanu2006,chow2011,chow2012}. Note that in order to achieve high-fidelity state preparation, additional flux bias can be added to the transmons, enabling fast tunability of the frequencies.
By suitably tuning the magnitude of the coupling capacitances and resistance, we set the coupling constant as $\mu=10^{-1.5}\hbar\omega_1\approx h\times0.16$ GHz. Taking into account the values that are usually measured in real experiments with transmon qubits \cite{Burnett2019}, we choose the local relaxation time $T_1=3\times 10^5/\omega_1\approx 10\,\mu s$. 
Finally, we set the inverse temperature $\beta$ such that $\hbar\omega_1\beta=10$, i.e. $\beta\approx 3\times 10^{24}J^{-1}$. This is reasonable, considering that it corresponds to a temperature $T\approx 24$ mK. This choice allows us to neglect the leakages to the third level of the transmon qubit, that would be proportional, according to Maxwell-Boltzmann distribution, to a very small factor of the order of $\approx 1/e^{20}$. Higher temperatures (as in some parameter regions of Figs.~\ref{fig:detVSmuBeta}(e)-(h) and~\ref{fig:further}(e)-(l)), however, may induce some non-negligible leakages to higher energy levels. In this case, cavity filters at the input of the two qubits can be installed to reduce the effect of spurious excitations to frequencies other than the qubit frequencies \cite{Cattaneo2021} -  see for instance the scheme in Ref.~\cite{ronzani2018tunable}, where a couple of resonant LC circuits are employed to suitably filter the radiation coming from separate thermal baths.

We can now investigate the open dynamics of the transmon qubits under the action of the resistor, with the parameters set as above, for the two different initial states associated to the figures of merit of synchronization and subradiance, so as to be able to track the evolution of respectively $\sigma_{1,2}^x(t)$ and $P_{1,2}^e(t)$. The complete information about these observables is acquired by local measurements on each qubit using the standard dispersive scheme. In this scheme, $P_{1,2}^e(t)$ are obtained directly as populations of the excited state, while for measuring $\sigma_{1,2}^x(t)$ we need to apply an $X$-pulse before the measurement. After performing these measurements, we can compute the synchronization and subradiance measures $Syn$ and $Sub$ as discussed in Sec.~\ref{sec:figuresOfM}. The results of our simulation are plotted in Fig.~\ref{fig:evolution}, where we also compare them with the negativity and collectiveness as a function of time (subfig.~(c)). The Pearson coefficient depicted in Fig.~\ref{fig:evolution}(a) shows the appearance of an antisynchronized regime (upper inset) after an incoherent transient (lower inset). Still, it does not correctly detect the synchronization time $t_S$ defined in Eq.~\eqref{eqn:syncTime}, since it provides a value higher than $99\%$ for $t\approx 4.2\times 10^2/\omega_1$. On the contrary, a more accurate spectral analysis shows that the time at which the slowest decaying mode in the dynamics of $\langle\sigma_{1,2}^x(t)\rangle$ is $100-$times bigger than all the other ones is $t_S\approx 8.3\times 10^2/\omega_1$, highlighting how the Pearson coefficient alone is not enough accurate to extract the value of $Syn$. Fig.~\ref{fig:evolution}(b) depicts the decay of $\langle P_{1,2}^e(t)\rangle$ as a function of time. We observe the emergence of a subradiant mode which remains alive also at late times, and a spectral analysis reveals the subradiance time Eq.~\eqref{eqn:subTime} $t_B\approx 3.4\times 10^2/\omega_1$. From this analysis, we extract the value of the figures of merit $Syn\approx 428$, $Sub\approx 529$, according to Eqs.~\eqref{eqn:syncMeasure} and~\eqref{eqn:syncSub}. Finally, let us focus on the negativity and collectiveness in Fig.~\ref{fig:evolution}(c): $\mathcal{N}(t)$ increases at $t\rightarrow 0^+$, showing that entanglement is generated as soon as the dynamics begins, as predicted in Appendix~\ref{sec:entProof}. After reaching the maximum value at $t\approx 21/\omega_1$, it decreases towards zero. Then, a less intense re-birth of entanglement is observed around $t\approx 50/\omega_1$, as already predicted in similar models \cite{Cattaneo2019}. The collectiveness displays a similar behavior, reaching the top at later times $t\approx 25/\omega_1$, and then decreasing with an oscillating behavior. Consistently with the original paper \cite{Rivas2015}, it decays toward $0$ at infinite time. From the evolution we obtain the values $\mathcal{N}_M\approx 0.37$ and $\bar{I}_M\approx 0.81$. It is important to note that all these characteristic times are smaller than the relaxation time $T_{1}$; this gives a realistic temporal window for performing measurements and observing these effects.

\section{Conclusions}
\label{sec:conclusions}

In this work we have investigated the behavior of quantum synchronization, 
subradiance, entanglement, and collectiveness of the dynamics, in a broad 
scenario of two superconducting transmon qubits dissipating into a common 
thermal bath. Although the four phenomena are essentially different, and a 
general one-to-one correspondence between them cannot be established, they share 
a common cause, namely the action of the collective bath. Inspired by this, 
we have addressed in which scenarios synchronization and subradiance (equipped 
with well-defined, time-independent quantifiers) can act as signatures of 
entanglement generation. We also have  compared  synchronization and subradiance 
with a measure of the \textit{collectiveness} of the evolution, i.e. of how 
much the dynamics differs from one described by two separable quantum maps 
acting locally on each qubit. Finally, we have  provided 
an experimental proposal   for a system of two transmon qubits coupled to the 
same resistor, which mimics a thermal bath.

We have found that quantum synchronization acts as a good signature of 
entanglement generation, i.e. the behavior of its measure  varies in the same 
way of the measure of entanglement production as a function of the model 
parameters, except in the scenario with the qubits coupled to the bath in an 
unbalanced way, which favors the emergence of a slowly-decaying eigenmode, while 
it decreases the collectiveness of the dynamics. The measure of subradiance  
follows the same behavior, but it fails to describe entanglement generation for 
high temperatures, where the slowly-decaying subradiant eigenmode is not 
affected by temperature, while the decay rate of all the other modes are 
enhanced. It is interesting to notice that, in the scenarios where synchronization and subradiance cannot be considered as reliable signatures of entanglement generation, the common bath can easily induce a certain collective phenomenon while struggles to produce a different one. Even more curiously, there are cases for which we may experimental detect intense subradiance and quantum synchronization (e.g. for very strongly unbalanced weights, see Figs.~\ref{fig:detVSg} and ~\ref{fig:further}(i)-(l)), which are clearly collective phenomena, while the measure of collectiveness is very weak. 

Our results also shed new light on the relation between entanglement generation 
and collectiveness. We have observed that, at least if the initial state of the 
system is symmetric with respect to the switching of the two qubits, their 
values follow exactly the same behavior, with the only exception that the 
collectiveness decreases but does not reach a null value for 
$T\rightarrow\infty$, while entanglement does. Our findings seem to suggest that 
the two measures are closely related, and a tighter connection between them may 
be found. Finally, our analysis establishes quantitatively the correlated dissipation 
strength needed for a pair of transmon qubits to display effects such as entanglement and spontaneous synchronization.

Remarkably, the advantage of employing synchronization or subradiance as 
signatures of entanglement generation consists in the fact that their measures 
can be computed by monitoring the mean value as a function of time of local 
observables only, without relying on joint measurements, which are on the 
contrary necessary to calculate any measure of entanglement, such as the 
negativity. We expect this to be especially useful when the system tomography is 
not available. In particular, it would be interesting to extend our results to 
systems of multiple qubits that realize a quantum computing platform, for which 
tomography is not feasible anymore. Furthermore, the experimental proposal we discuss may be implemented with the currently available technology and know-how about superconducting qubits. Its realization would be the first demonstration of spontaneous quantum synchronization, which does not require the presence of an external driving field, in the deep quantum regime.

\medskip
\textbf{Acknowledgements} \par
The numerical analyses have been performed with the assistance of QuTiP, the Quantum Toolbox in Python \cite{Johansson2013}. This project has received funding from the European Union’s Horizon 2020 research and innovation programme under grant agreement No. 862644 (FET Open--QUARTET), from the Spanish State Research Agency through the QUARESC project (PID2019-109094GB-C21 /AEI/10.13039/501100011033) and the project QUAREC funded by CAIB. S.M. acknowledges financial support from the Academy of Finland via the Centre of Excellence program (Project no. 336814). MC acknowledges funding from the
Severo Ochoa and María de Maeztu Program for Centers and Units of Excellence in
R\&D (MDM-2017-0711), from Fondazione Angelo della Riccia and from the  Finnish Center of Excellence in Quantum Technology
QTF (Academy of Finland project 312296). G.L.G. is funded by the Spanish Ministerio de Educación y Formación Profesional/Ministerio de Universidades and co-funded by the University of the Balearic Islands through the Beatriz Galindo program (BG20 /00085). G. S. P. acknowledges financial support from the Academy of Finland RADDESS project 328193.

\appendix
\section{Coefficients of the master equation}
\label{sec:coefficients}
We assume that the bath is in a thermal state at temperature $T$. The coefficients of the master equation~\eqref{eqn:masterEqLiouvillian} read:
\begin{equation}
\label{eqn:coefficients}
\begin{split}
\gamma_{jk}^\downarrow&=g_jg_k(\Gamma_\beta(\omega_j)+\Gamma_\beta(\omega_k)^*)+1/T_1\,\delta_{jk},\\
\gamma_{jk}^\uparrow&=g_jg_k(\Gamma_\beta(-\omega_j)+\Gamma_\beta(-\omega_k)^*),\\
s_{jk}^\downarrow&=g_jg_k(\Gamma_\beta(\omega_j)-\Gamma_\beta(\omega_k)^*)/2i,\\
s_{jk}^\uparrow&=g_jg_k(\Gamma_\beta(-\omega_j)-\Gamma_\beta(-\omega_k)^*)/2i,\\
\end{split}
\end{equation}
where $\Gamma_\beta(\omega)$ is the one-side Fourier transform of the autocorrelation functions of the bath, which depends on the spectral density Eq.~\eqref{eqn:spectralDensity} and on the temperature of the thermal bath through $\beta=1/k_B T$ \cite{BreuerPetruccione,Cattaneo2019}:
\begin{equation}
\label{eqn:Gamma}
\begin{split}
\Gamma_\beta(\omega)=&\pi(N_\beta(\omega)+1)J(\omega)\\
&+i\mathcal{P}\int_0^\infty d\omega_k' J(\omega_k') \left(\frac{N_\beta(\omega_k')+1}{\omega-\omega_k'}+\frac{N_\beta(\omega_k')}{\omega+\omega_k'}\right),
\end{split}
\end{equation}
where $N_\beta(\omega)=\frac{1}{e^{\beta\hbar\omega}-1}=\frac{1}{2}\left(\coth\frac{\beta\hbar\omega}{2}-1\right)$, and $J(\omega)$ is the spectral density of the bath, defined as:
\begin{equation}
\label{eqn:spectralDensity}
J(\omega)=\frac{\mu^2}{\hbar^2}\sum_k f_k^2 \delta(\omega-\Omega_k),
\end{equation} 
where $f_k$ are introduced in Eq.~\eqref{eqn:HamiltonianTotal}.
In the analysis of Sec.~\ref{sec:results} we have employed an Ohmic spectral density, i.e.:
\begin{equation}
\label{eqn:ohmic}
J(\omega)=\frac{\mu^2}{\hbar^2\omega_1^2}\frac{\omega\omega_C^2}{\omega_C^2+\omega^2},
\end{equation}
where $\omega_C$ is a cut-off frequency that we have set as $\omega_C=20\omega_1$, and we have chosen to renormalize the spectral density by the frequency of the first qubit.

In the decay coefficients of Eq.~\eqref{eqn:coefficients} we have added a local dissipative decay rate $1/T_1$ on each qubit, which represents the action of a phenomenological local bath at zero temperature. Its effects are therefore not relevant for the absorption coefficients, and we have also neglected its contribution to the Lamb-shift. These choices are motivated by the experimental decay observed in a single superconducting qubit, which reaches to a good approximation the ground state at infinite time, corresponding to a thermal bath at zero temperature. On the contrary, we allow for the common bath to have a finite temperature, which experimentally can be realized by local heating of the resistor connecting the qubits. 

\section{Blocks of the Liouvillian superoperator}
\label{sec:Liouvillian}
The master equation~\eqref{eqn:masterEqLiouvillian} is in partial secular approximation and therefore a fundamental symmetry on the superoperator level emerges \cite{Cattaneo2020}: $[\mathcal{L},\mathcal{N}]=0$, where $\mathcal{N}=[P_1^e+P_2^e,\cdot\,]$ is the \textit{number superoperator} ($P_1^e$ and $P_2^e$ are defined in Eq.\eqref{eqn:emissionObs}). For this reason, the Liouvillian superoperator can be block-diagonalized with each block labeled by a different eigenvalue of $\mathcal{N}$. Let us work in the basis of the space of operators $\{\ket{jk}\bra{lm}\}_{j,k,l,m=e,g}$. $\mathcal{N}$ is diagonal in this basis, and its eigenvalues  correspond to the number of excited qubit states in the ket minus the number of excited qubit states in the bra. Therefore, a single eigenvalue $d$ can assume the values $d=-2,-1,0,1,2$, and the Liouvillian superoperator is divided into five independent blocks labeled by $d$: $\mathcal{L}=\bigoplus_{d=-2}^{2}\mathcal{L}_d$. In particular, the block $\mathcal{L}_0$ will act on the basis vectors $\ket{ee}\bra{ee}$, $\ket{eg}\bra{eg},\ket{eg}\bra{ge},\ket{ge}\bra{eg},\ket{eg}\bra{eg}$,\\$\ket{gg}\bra{gg}$, $\mathcal{L}_1$ on $\ket{ee}\bra{eg},\ket{ee}\bra{ge},\ket{eg}\bra{gg},\ket{ge}\bra{gg}$, $\mathcal{L}_{-1}$ on $\ket{eg}\bra{ee},\ket{ge}\bra{ee},\ket{gg}\bra{eg},\ket{gg}\bra{ge}$, $\mathcal{L}_2$ on $\ket{ee}\bra{gg}$ and $\mathcal{L}_{-2}$ on $\ket{gg}\bra{ee}$. In these bases, we have $\mathcal{L}_{-1}=\mathcal{L}_1^*$ and $\mathcal{L}_{-2}=\mathcal{L}_2^*$ (more details in Refs.~\cite{Bellomo2017,Cattaneo2020}). The spectral analysis of each block of the Liouvillian superoperator leads to the solution of the dynamics as given by Eq.~\eqref{eqn:evstate}.

\section{Derivation of the figures of merit}
\label{sec:derFig}
\subsection{Quantum synchronization}
The Pearson coefficient is defined in a temporal window $\Delta t$ as:
\begin{equation}
\label{eqn:pearson}
\mathcal{C}_{\Delta  t}(t)=\frac{\int_t^{t+\Delta t} (\sigma_1^x(t')-\bar{\sigma_1}^x)(\sigma_2^x(t')-\bar{\sigma_2}^x)dt'}{\sqrt{\prod_{k=1}^2\int_t^{t+\Delta t} (\sigma_k^x(t')-\bar{\sigma_k}^x)^2}dt'},
\end{equation}
where $\bar{\sigma_k}^x=\frac{1}{\Delta t}\int_t^{t+\Delta t} \sigma_k^x(t') dt'$. In a scenario in which the dynamics of the observables reaches the perfect in-phase synchronization at a given time $t_S$, we observe that the value of the Pearson coefficient stabilizes to $1$ at the same time, for a suitable time window $\Delta t$ \cite{Giorgi2012,Giorgi2013,Karpat2020}. A slightly different definition of the Pearson coefficient allows us to catch the emergence of synchronization with a given phase shift \cite{Galve2016}.

Let us now focus on a time-independent measure of quantum synchronization. It will depend only on the separation between the real part of the eigenvalues of the modes of $\mathcal{L}_1$, on each corresponding inner product $c_{jk}^{(1)}$ and on the initial conditions, i.e. on the state at which we initialize the system at $t=0$. We start the dynamics from the state $\rho_S(0)=\ket{\psi_{Syn}}\bra{\psi_{Syn}}$ defined in Sec.~\ref{sec:sync}, and we monitor the observables $\sigma_1^x(t)$ and $\sigma_2^x(t)$. The free evolution of each qubit would give $\langle\sigma_k^x(t)\rangle=\cos(\omega_k t+\phi_k)$ where $\phi_k$ is a phase depending on the initial conditions, while the presence of a common bath can lead to the synchronization of their dynamics, i.e. to a situation in which they oscillate at the same frequency. To do so, their evolution must be monochromatic, that is to say, the main contribution to the mean value of $\sigma_k^x(t)$ in Eq.~\eqref{eqn:sigmaX} must be given by a single eigenmode of the Liouvillian block $\mathcal{L}_1$ only. This happens when all the other modes have disappeared due to their faster decay. Let us say that synchronization emerges at the time $t_S$, when the contribution of the slowest-decay mode with eigenvalue $\lambda_4^{(1)}$ to $\langle\sigma_1^x(t)\rangle$ and $\langle\sigma_2^x(t)\rangle$ is 100-times bigger than the one of any other mode. Intuitively, at $t_S$ the Pearson coefficient will have reached a value $\mathcal{C}_{\Delta t}(t_S)> 0.99$. Using the notation introduced in Eq.~\eqref{eqn:sigmaX}, the synchronization time $t_S$ is estimated by inverting
\[
 \abs{p_{04}^{(1)} c_{4k}^{(1)}}e^{\Re(\lambda_4^{(1)})t_S}=100\cdot \abs{p_{0j}^{(1)} c_{jk}^{(1)}}e^{\Re(\lambda_j^{(1)})t_S},
\]
and then maximazing over $j=2,3,4$ and $k=1,2$. Finally, we get:
\begin{equation}
\label{eqn:syncTime}
t_S=\max_{j=1,2,3;k=1,2}\frac{\log(\frac{100\abs{p_{0j}^{(1)} c_{jk}^{(1)}}}{\abs{p_{04}^{(1)} c_{4k}^{(1)}}})}
{\Re(\lambda_4^{(1)})-\Re(\lambda_j^{(1)})}.
\end{equation}
We observe that the synchronization time depends on the separation between the slowest decay rate and all the other ones, and on the inner product between the eigenvector of the slowest-decaying mode and the initial state, $\sigma_1^x(t)$ and $\sigma_2^x(t)$.

If the two slowest decay rates do not coincide, there will always be a finite synchronization time at which the mean values of the qubit observables will oscillate at the same frequency. However, decoherence appears along the evolution and the synchronization time may be way longer than the time at which the system has lost all of its coherences, and the detection of synchronization would be almost impossible. To take into account both phenomena, we define the figure of merit of synchronization as the ratio between the synchronization time and the time at which the slowest-decaying mode is reduced to $1/100$ of its initial value, and we obtain Eq.~\eqref{eqn:syncMeasure}:
\begin{equation}
\label{eqn:syncMeasureBis}
Syn=\abs{\frac{\log 100}{t_{S}\Re(\lambda_4^{(1)})}}.
\end{equation}

Experimentally, $Syn$ can be estimated by monitoring the mean value of $\sigma_1^x(t)$ and $\sigma_2^x(t)$ until the time at which the spectral density of the signal reveals the presence of a single mode only, with an error of the $1\%$. Then, the decay rate of the remaining mode can be found through an exponential fit of the carrier wave. The higher $Syn$, the easier will be to detect quantum synchronization, since the signals of the mean value of $\sigma_1^x(t)$ and $\sigma_2^x(t)$ will be more intense at the synchronization time. If $Syn=1$, then the synchronization appears exactly when the only remaining mode is reduced to $1/100$ of its initial value.

\subsection{Subradiance}
The derivation of the figure of merit for subradiance is analogous to the one of quantum synchronization: we want to identify a collective slowly-decaying mode of the system decay which remains alive when all the other modes have vanished. The only differences consist in the observables we want to monitor, that now are $P_1^e(t)$ and $P_2^e(t)$ instead of $\sigma_1^x(t)$ and $\sigma_2^x(t)$, and in the initial conditions, that read $\rho_S(0)=\frac{(\ket{eg}-\ket{ge})(\bra{eg}-\bra{ge})}{2}$.

In a scenario without a collective bath acting on both the qubits, each of them would decay with an independent decay rate. On the contrary, the presence of a common bath creates a slowly-decaying collective mode whose component in both $P_1^e(t)$ and $P_2^e(t)$ survives after all the other ones have vanished, apart from the mode pertaining to the steady state which does not decay at all (say the one with eigenvalue $\lambda_6^{(0)}=0$). We want to identify the long-surviving component, let us say corresponding to the eigenvalue $\lambda_5^{(0)}$, and the time at which it emerges. Following the discussion about quantum synchronization, we choose the subradiance time $t_B$ as the time at which the slowest-decaying component in the mean value of $P_1^e(t_B)$ and $P_2^e(t_B)$ is 100-times bigger than that of any other mode (apart from the steady state one). Using Eq.\eqref{eqn:sigmaZ} and analogously to Eq.\eqref{eqn:syncTime}, we have:
\begin{equation}
\label{eqn:subTime}
t_B=\max_{j=1,2,3,4;k=1,2}\frac{\log(\frac{100\abs{p_{0j}^{(0)} h_{jk}^{(0)}}}{\abs{p_{05}^{(0)} h_{5k}^{(0)}}})}
{\Re(\lambda_5^{(0)})-\Re(\lambda_j^{(0)})}.
\end{equation}
As for the case of quantum synchronization, we define the figure of merit of subradiance as the ratio between the subradiance time $t_B$ and the time at which the component of the slowest-decaying mode $\lambda_5^{(0)}$ is reduced to $1/100$ of its initial value, and we obtain Eq.~\eqref{eqn:syncSub}.

Analogously to the estimation of the figure of merit of quantum synchronization, in order to compute the value of the subradiance measure Eq.~\eqref{eqn:syncSub} in a real experiment one has to track the signals of the mean values $\langle P_1^e(t)\rangle $ and $\langle P_2^e(t)\rangle$. In this way, we are able to detect the time $t_B$ at which one single decaying component of the signals is 100-times bigger than any other one (after having removed the steady-state one), and to estimate the corresponding decay rate.

\section{Sufficient conditions for the entangling power of the bath}
\label{sec:entProof}
We follow the discussion in Ref.~\cite{Benatti2008} in order to provide some sufficient conditions assuring that the master equation~\eqref{eqn:masterEqLiouvillian} has the capability of producing entanglement for some values of its coefficients. $\mathcal{T}_2$ represents the partial transpose with respect to the qubit $2$. According to the partial transpose criterion \cite{PhysRevA.65.032314}, a two-qubit state $\rho$ is entangled if and only if $\tilde{\rho}=\mathcal{T}_2[\rho]$ is negative-definite. Let us define $\tilde{\mathcal{L}}=\mathcal{T}_2\circ \mathcal{L}\circ\mathcal{T}_2$, then we observe that $\tilde{\mathcal{L}}[\tilde{\rho}]=\mathcal{T}_2[\mathcal{L}[\rho]]$. It can be shown \cite{Benatti2008} that a semigroup driven by $\mathcal{L}$ is entangling if there exist a separable pure state $\rho$ and a vector $\ket{\Phi}\in\mathbb{C}_4$ such that $\mel{\Phi}{\tilde{\rho}}{\Phi}=0$ and $\mel{\Phi}{\tilde{\mathcal{L}}[\tilde{\rho}]}{\Phi}<0$.

Noticing that $\mathcal{T}_2[A\otimes B \rho]=A\otimes\mathbb{I}\,\mathcal{T}_2[\rho]\,\mathbb{I}\otimes B^T$ and $\mathcal{T}_2[ \rho A\otimes B]=\mathbb{I}\otimes B^T\,\mathcal{T}_2[\rho]\,A\otimes\mathbb{I}$, the form of $\tilde{\mathcal{L}}$ can be derived from Eq.~\eqref{eqn:masterEqLiouvillian} and reads:
\begin{equation}
\label{eqn:LiouvillianTilde}
\begin{split}
\tilde{\mathcal{L}}[\rho]=&-\frac{i}{2}[(\omega_1+s_{11})\sigma_1^z-(\omega_2+s_{22})\sigma_2^z,\rho]-i({s}_{+}(\sigma_1^-\rho\sigma_2^--\sigma_2^-\rho\sigma_1^-)+{s}_{-}(\sigma_1^+\rho\sigma_2^+-\sigma_2^+\rho\sigma_1^+))\\
&+\sum_{j=1,2}\left[\gamma_{jj}^\downarrow\left(\sigma_j^-\rho\sigma_j^+-\frac{1}{2}\{\rho,\sigma_j^+\sigma_j^-\}  \right)+\gamma_{jj}^\uparrow\left(\sigma_j^+\rho\sigma_j^--\frac{1}{2}\{\rho,\sigma_j^-\sigma_j^+\}  \right)\right]+\gamma_{12}^\downarrow\sigma_1^-\sigma_2^-\rho+\gamma_{12}^\uparrow\sigma_1^+\sigma_2^+\rho\\
&-\frac{\gamma_{12}^\downarrow+\gamma_{21}^\uparrow}{2}(\sigma_1^-\rho\sigma_2^-+\sigma_2^-\rho\sigma_1^-)-\frac{\gamma_{21}^\downarrow+\gamma_{12}^\uparrow}{2}(\sigma_1^+\rho\sigma_2^++\sigma_2^+\rho\sigma_1^+)+\gamma_{21}^\downarrow\rho\sigma_1^+\sigma_2^++\gamma_{21}^\uparrow\rho\sigma_1^-\sigma_2^-,
\end{split}
\end{equation}
with $s_{jj}=s_{jj}^\downarrow-s_{jj}^\uparrow$, ${s}_{+}=s_{12}^\downarrow+s_{21}^\uparrow$, ${s}_{-}=s_{12}^\uparrow+s_{21}^\downarrow$. 

We write the separable state $\rho$ as $\rho=\ket{\psi}\bra{\psi}\otimes\ket{\varphi}\bra{\varphi}$. Then, $\tilde{\rho}=\ket{\psi}\bra{\psi}\otimes\ket{\varphi^*}\bra{\varphi^*}$, where $\ket{\varphi^*}$ is the state whose components in the canonical basis of $\sigma_2^z$ are the complex conjugate of the ones of $\ket{\varphi}$. We now have to find a suitable vector $\ket{\Phi}\in\mathbb{C}_4$ with $\mel{\Phi}{\tilde{\rho}}{\Phi}=0$. The latter condition tells us that we need to choose it in the orthogonal component of the subspace generated by the state $\ket{\Psi_1}=\ket{\psi}\otimes\ket{\varphi^*}$, and a suitable basis is expressed as $\ket{\Psi_2}=\ket{\psi}\otimes\ket{\varphi^*_\bot}$, $\ket{\Psi_3}=\ket{\psi_\bot}\otimes\ket{\varphi^*}$, $\ket{\Psi_4}=\ket{\psi_\bot}\otimes\ket{\varphi^*_\bot}$. Let us define the matrix $M_{ij}=\mel{\Psi_i}{\tilde{\mathcal{L}}[\tilde{\rho}]}{\Psi_j}$. Then, it is sufficient \cite{Benatti2008} that $M_{22} M_{33}<\abs{M_{23}}^2$ for the bath to create entanglement between the qubits at time $t\rightarrow 0^+$, starting from $\rho_S(0)=\rho$. We now consider different initial conditions and discuss their consequences:
\begin{itemize}
\item If $\rho_S(0)=\ket{gg}\bra{gg}$ or $\rho_S(0)=\ket{ee}\bra{ee}$, we have $M_{23}=0$ and $M_{22},M_{33}> 0$ for any $T>0$, therefore $\lim_{t\rightarrow 0^+}\rho_S(t)$ would still be separable in any scenario.
\item If $\rho_S(0)=\ket{ge}\bra{ge}$, the sufficient condition for generating entanglement at time $t\rightarrow 0^+$ reads $\gamma_{22}^\downarrow\gamma_{11}^\uparrow<\left(\frac{\gamma_{12}^\downarrow+\gamma_{21}^\uparrow}{2}\right)^2+s_+^2$. In particular, we see that the bath always generates entanglement if $T=0$. In any case, we observe that the quantity $s_+^2$ is relevant to provide a general sufficient condition for the bath to be entangling. An analogous result is found for $\rho_S(0)=\ket{eg}\bra{eg}$.
\item If $\rho_S(0)=\rho_{C}=1/4\sum_{j,k,l,m=e,g}\ket{jk}\bra{lm}$, i.e. the state we choose as initial state of the evolution in our analysis, as discussed in Sec.~\ref{sec:entanglement}, the condition is $(\gamma_{22}^\downarrow+\gamma_{22}^\uparrow)(\gamma_{11}^\downarrow+\gamma_{11}^\uparrow)<\left(\frac{\gamma_{12}^\downarrow+\gamma_{12}^\uparrow+\gamma_{21}^\downarrow+\gamma_{21}^\uparrow}{2}\right)^2+(s_++s_-)^2$. Once again, we recognize the important role the Lamb-shift terms $s_\pm$ play in assuring a sufficient condition to generate entanglement.
\end{itemize}
Note that, a priori, if for an initial state we find that entanglement is not generated at time $t\rightarrow 0^+$, this does not mean that entanglement will never appear during its evolution. Indeed, it may be the case that, after a certain time $t^*$, the state is represented by a density matrix $\rho$ for which the condition $M_{22} M_{33}<\abs{M_{23}}^2$ is verified. In other words, the method we have followed \cite{Benatti2008} is useful to provide sufficient conditions, while in order to find a necessary one we would need to check that $M_{22} M_{33}>\abs{M_{23}}^2$ for any initial separable pure state. Furthermore, this procedure does not give any information about the strength of the generated entanglement. To compute it, we need to study the evolution of the state and to find the maximum value of the negativity (see Eq.~\eqref{eqn:negMeasure}), as we have done in Sec.~\ref{sec:results}.

\section{Discussion on the analytical limits}
\label{sec:an}
\subsection{$T=0$}
In the zero temperature case, we can derive analytically the eigenvalues and some of the eigenvectors of the Liouvillian superoperator $\mathcal{L}$, as already discussed in Ref.~\cite{Bellomo2017}. In particular, the eigenvalues of the block $\mathcal{L}_0$ read:
\begin{equation}
\begin{split}
\lambda_1^{(0)}&=-\gamma_1-\gamma_2,\\
\lambda_2^{(0)}&=-\frac{\gamma_1+\gamma_2}{2}-\Re(V),\\
\lambda_3^{(0)}&=-\frac{\gamma_1+\gamma_2}{2}+i\Im(V),\\
\lambda_4^{(0)}&=-\frac{\gamma_1+\gamma_2}{2}-i\Im(V),\\
\lambda_5^{(0)}&=-\frac{\gamma_1+\gamma_2}{2}+\Re(V),\\
\lambda_6^{(0)}&=0,\\
\end{split}
\end{equation}
where
\begin{equation}
\begin{split}
V=\sqrt{(\gamma_{12}^\downarrow+2i s_{12}^\downarrow)(\gamma_{21}^\downarrow+2i s_{21}^\downarrow)+\left(\frac{\gamma_{11}^\downarrow-\gamma_{22}^\downarrow}{2}+ i\Delta \omega\right)^2},
\end{split}
\end{equation}
taking the principal square root.
Note that we have sorted the eigenvalues in descending order of absolute value of their real part. The subradiant mode we are interested in has eigenvalue $\lambda_5^{(0)}$. The associated right eigenvector is $\tau_5^{(0)}=\ket{S_R}\bra{S_R}-\ket{gg}\bra{gg}$, where $
\ket{S_R}=\frac{\alpha_S\ket{eg}+\ket{ge}}{\sqrt{1+\abs{\alpha_S}^2}}$, with 
\[
\alpha_S=\frac{\frac{\gamma_{11}^\downarrow-\gamma_{22}^\downarrow}{2}+i\Delta\omega+V}{\sqrt{(\gamma_{12}^\downarrow+i s_{12}^\downarrow)(\gamma_{21}^\downarrow+i s_{21}^\downarrow)}}.
\]
Note that, according to Eq.~\eqref{eqn:syncTime}, we need to consider the difference between the real parts of $\lambda_5^{(0)}$ and the rest of eigenvalues of $\mathcal{L}_0$, excluding $\lambda_6^{(0)}$ because it indicates the steady state eigenmode. Therefore, the smallest separation between the eigenvalues is given by $\Re(\lambda_5^{(0)})-\Re(\lambda_4^{(0)})=\Re(V)$. 

Let us now focus on the block $\mathcal{L}_1$. Its eigenvalues are:
\begin{equation}
\begin{split}
\lambda_1^{(1)}&=-\frac{1}{2}(3\frac{\gamma_{11}^\downarrow+\gamma_{22}^\downarrow}{2}+V^*)-i\frac{\omega_1+\omega_2}{2},\\
\lambda_2^{(1)}&=-\frac{1}{2}(3\frac{\gamma_{11}^\downarrow+\gamma_{22}^\downarrow}{2}-V^*)-i\frac{\omega_1+\omega_2}{2},\\
\lambda_3^{(1)}&=-\frac{1}{2}(\frac{\gamma_{11}^\downarrow+\gamma_{22}^\downarrow}{2}+V)-i\frac{\omega_1+\omega_2}{2},\\
\lambda_4^{(1)}&=-\frac{1}{2}(\frac{\gamma_{11}^\downarrow+\gamma_{22}^\downarrow}{2}-V)-i\frac{\omega_1+\omega_2}{2},\\
\end{split}
\end{equation}
where once again we have sorted them in desceding order of absolute value of the real part. The right eigenvector associated to the slowest eigenvalue $\lambda_4^{(1)}$ is $\tau_4^{(1)}=\ket{A_R}\bra{gg}$, where $\ket{A_R}=\frac{\alpha_A\ket{eg}+\ket{ge}}{\sqrt{1+\abs{\alpha_A}^2}}$, and
\[
\alpha_A=\frac{\frac{\gamma_{11}^\downarrow-\gamma_{22}^\downarrow}{2}+i\Delta\omega-V}{\sqrt{(\gamma_{12}^\downarrow+i s_{12}^\downarrow)(\gamma_{21}^\downarrow+i s_{21}^\downarrow)}}.
\]
Once again, we find $\Re(\lambda_4^{(1)})-\Re(\lambda_3^{(1)})=\Re(V)$. Let us now calculate this quantity:
\begin{equation}
\label{eqn:reV}
\Re(V)=\sqrt{\frac{\sqrt{x^2+y^2}+x}{2}},
\end{equation}
with
\begin{equation}
\begin{split}
x&=\frac{(\gamma_{11}^\downarrow-\gamma_{22}^\downarrow)^2}{4}-\Delta\omega^2+\gamma_{12}^\downarrow\gamma_{21}^\downarrow-4s_{12}^\downarrow s_{21}^\downarrow,\\
y&=4\gamma_{21}^\downarrow s_{12}^\downarrow+4 \gamma_{12}^\downarrow s_{21}^\downarrow+\Delta\omega (\gamma_{11}^\downarrow-\gamma_{22}^\downarrow) .\\
\end{split}
\end{equation}
As discussed in the main text, by assuming $g_1=g_2$ and consequently $\gamma_{11}^\downarrow-\gamma_{22}^\downarrow\approx 0$, clearly we have $\frac{d}{d\Delta\omega}\Re(V)<0$. 

Finally, according to the definition in Eq.~\eqref{eqn:sigmaZ}, if $\gamma_{11}^\downarrow-\gamma_{22}^\downarrow= 0$ and $\Delta\omega=0$, we have $h_{51}^{(0)}=\Tr[P_1^e \tau_5^{(0)}]=h_{52}^{(0)}=\Tr[P_2^e \tau_5^{(0)}]=\frac{1}{2}$. Analogously for synchronization and Eq.~\eqref{eqn:sigmaX}, we have $\abs{c_{41}^{(1)}}=\abs{\Tr[\sigma_1^x \tau_4^{(1)}]}=\abs{h_{42}^{(1)}}=\abs{\Tr[\sigma_2^x \tau_4^{(1)}]}=\frac{1}{2}$. This means that for the unbalanced scenario with $g_1=g_2$, the zero detuning configuration provides maximal symmetric weights for the definition of subradiance ($h_{5k}^{(0)}$) and synchronization ($c_{4k}^{(1)} $), therefore the smaller the detuning the higher $Syn$ and $Sub$.

\subsection{$T\rightarrow \infty$}
Let us consider the limit of $T\rightarrow\infty$, and for simplicity $g_1=g_2$, $\Delta\omega=0$. In this simplified scenario, the master equation depends on only three parameters: $\omega=\omega_1+s_{11}=\omega_2+s_{22}$, $\gamma=\gamma_{jk}^\downarrow=\gamma_{jk}^\uparrow$, $s_+=s_-=s_{12}^\downarrow+s_{21}^\uparrow$, where $s_{jj}=s_{jj}^\downarrow-s_{jj}^\uparrow$ is a local Lamb-shift correction. Clearly, the measure of subradiance $Sub$ always takes an infinite value, since, as discussed in the main text, the symmetric scenario makes a decoherence-free subspace emerge, associated to the subradiant mode. Let us now focus on synchronization. The eigenvalues of the block $\mathcal{L}_1$ are the following:
\begin{equation}
\begin{split}
\lambda_1^{(1)}&=-3\gamma-\sqrt{4\gamma^2-s_+^2}-i\omega,\\
\lambda_2^{(1)}&=-3\gamma+\sqrt{4\gamma^2-s_+^2}-i\omega,\\
\lambda_3^{(1)}&=-\gamma-i(\omega-s_+),\\
\lambda_4^{(1)}&=-\gamma-i(\omega+s_+).\\
\end{split}
\end{equation}
We observe that the two slowest eigenvalues have the same real part, while different imaginary part. This means that synchronization can never appear ($Syn=0$), and after a transient time we will observe the presence of only two coexisting modes (corresponding to $\lambda_3^{(1)}$ and $\lambda_4^{(1)}$) oscillating with different frequencies. Therefore, we observe that the limit $T\rightarrow \infty$ in the balanced scenario increases the value of the measure of subradiance $Sub$, while it decreases the value of $Syn$, not allowing synchronization to emerge. This result is not valid anymore in the presence of unbalanced weights, i.e. $g_1\neq g_2$.

\begin{figure*}
\centering
\subfloat[]{%
  \includegraphics[scale=0.6]{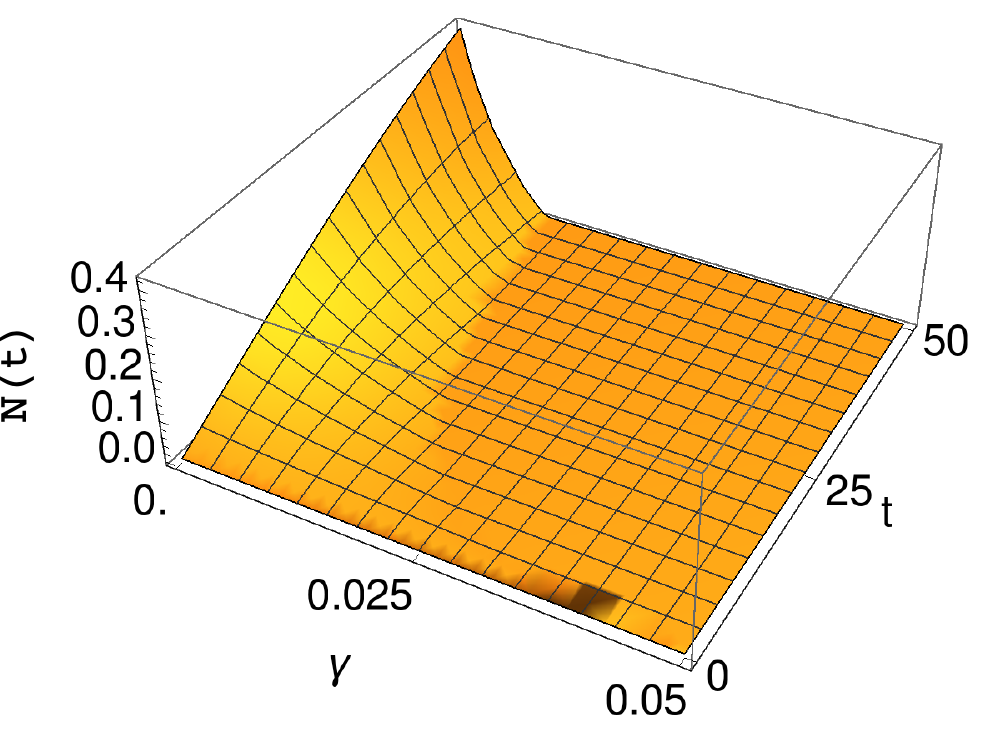}%
}\qquad\qquad
\subfloat[]{%
  \includegraphics[scale=0.6]{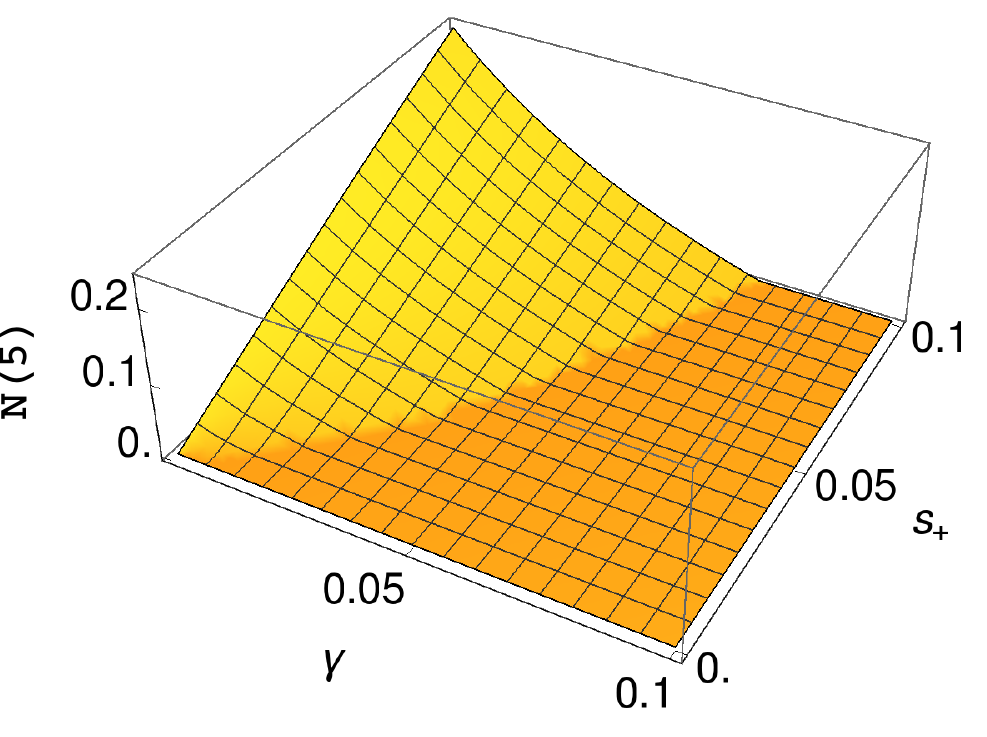}%
}\\
\subfloat[]{%
  \includegraphics[scale=0.6]{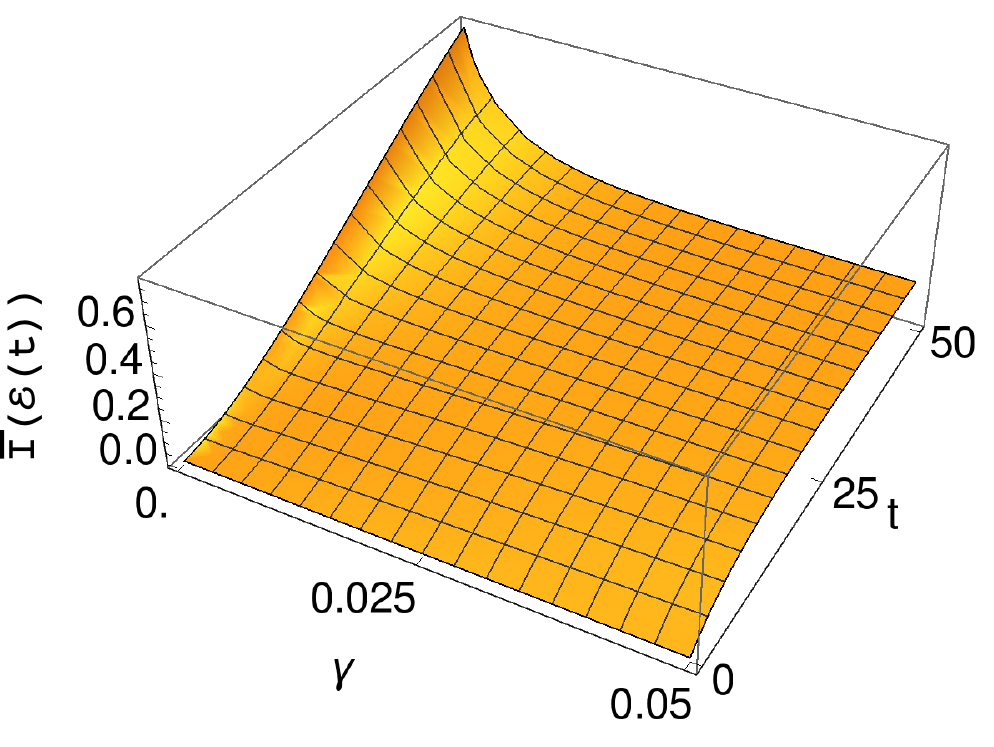}%
}\qquad\qquad
\subfloat[]{%
  \includegraphics[scale=0.6]{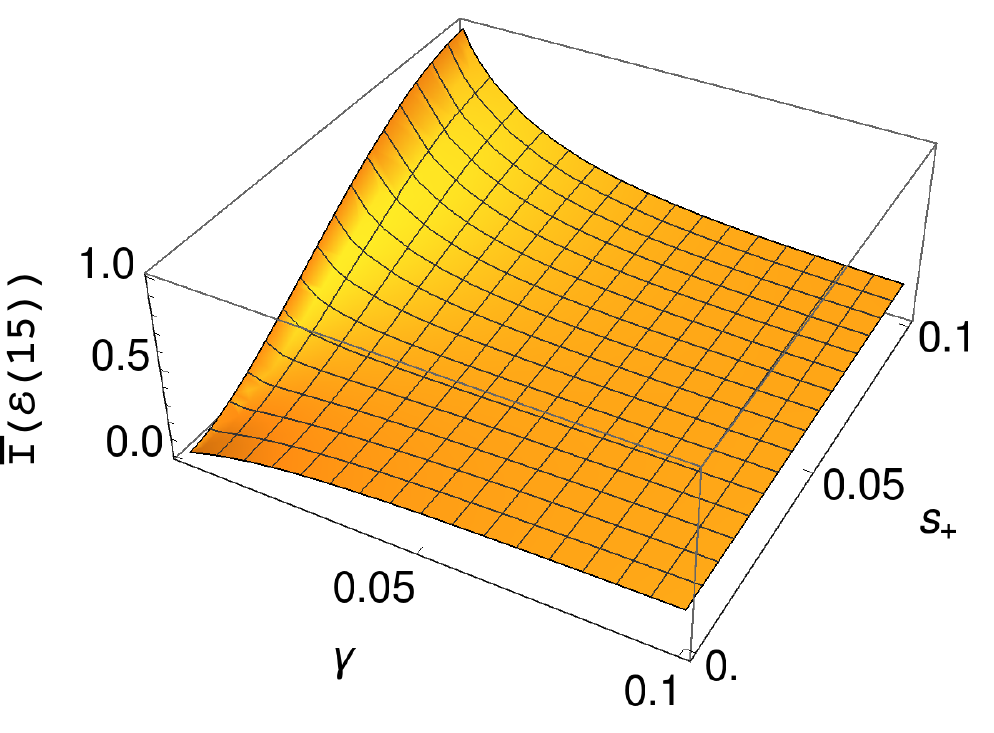}%
}\\
\caption{Entanglement (Figs.~(a)-(b)) and collectiveness (Figs.~(c)-(d)) of the quantum map generated by the master equation~\eqref{eqn:masterEqQmap}, as a function of time $t$, the dissipative interaction $\gamma$ and the collective Lamb-shift interaction $s_+$, having fixed $\hbar=\omega=1$. In particular, in Figs.~(a) and~(c) we have fixed $s_+=-0.02$ and varied $\gamma$ and $t$, in Fig.~(b) we have fixed $t=5$ and varied $\gamma$ and $s_+$, while in Fig.~(d) we have fixed $t=15$ and varied $\gamma$ and $s_+$.}
\label{fig:gvsS}
\end{figure*}

Let us now consider entanglement and collectiveness in the same scenario. Unfortunately, a readable analytical solution for the negativity $\mathcal{N}(t)$ in Eq.~\eqref{eqn:negativity} and collectiveness of the dynamics up to time $t$, $\bar{I}(\mathcal{E}(t))$ in Eq.~\eqref{eqn:measureCorr}, is not available. However, we can understand their behavior by looking at the ratio between the Lamb-shift collective term $s_+$ and the dissipative collective term $\gamma$. To do so, we analyze the behavior of the following master equation:
\begin{equation}
\label{eqn:masterEqQmap}
\frac{d}{dt}\rho_S(t)=\mathcal{L}_{\infty}[\rho_S(t)],
\end{equation}
with Liouvillian
\begin{equation}
\label{eqn:Liouvillian}
\begin{split}
\mathcal{L}_{\infty}[\rho]=&-i\left[\frac{\omega}{2}(\sigma_1^z+\sigma_2^z)+s_+(\sigma_1^+\sigma_2^-+\sigma_1^-\sigma_2^+),\rho\right]\\
&+\gamma\sum_{j,k=1,2}\left(\sigma_j^-\rho\sigma_k^+-\frac{1}{2}\{\sigma_k^+\sigma_j^-,\rho\}\right)\\
&+\gamma\sum_{j,k=1,2}\left(\sigma_j^+\rho\sigma_k^--\frac{1}{2}\{\sigma_k^-\sigma_j^+,\rho\}\right),
\end{split}
\end{equation}
corresponding to a scenario with a thermal bath at infinite temperature, but keeping the possibility to freely set the parameters $\omega$, $\gamma$ and $s_+$, which now do not depend on the spectral density of a specific bath. To work with dimensionsless units, we set $\hbar=1$ and $\omega=1$, so that we only need to focus on $\gamma$ and $s_+$.

In Fig.~\ref{fig:gvsS} we depict the entanglement and collectiveness as a function of $\gamma$ and $s_+$. In Figs.~\ref{fig:gvsS}(a),(c) we fix $s_+$ and make the time vary: we observe that, for fixed $s_+$, higher values of $\gamma$ lead to lower values of both entanglement and collectiveness. In Figs.~\ref{fig:gvsS}(b),(d) we fix a early time of the evolution, so as to compute the figures of merit in the first rising oscillation (compare with the simulated curve in Fig.~\ref{fig:evolution}(c)), and vary $\gamma$ and $s_+$. We clearly identify a trade-off between these two parameters: the higher the ratio $s_+/\gamma$, the higher the values of entanglement and collectiveness. These results confirm our claims in Sec.~\ref{sec:anL} of the main text.

\bibliography{syncSignature}

\end{document}